# Amplification and ellipticity enhancement of sub-femtosecond XUV pulses in IR-field-dressed neon-like active medium of a plasma-based X-ray laser

V.A. Antonov[1,*], I.R. Khairulin[1], M.Yu. Ryabikin[1,2], M.A. Berrill[3], V.N. Shlyaptsev[3], J.J. Rocca[3,4], and Olga Kocharovskaya[5]

[1]*Institute of Applied Physics of the Russian Academy of Sciences,*
*46 Ulyanov Street, Nizhny Novgorod 603950, Russia*

[2]*Lobachevsky State University of Nizhny Novgorod,*
*23 Prospekt Gagarina, Nizhny Novgorod 603950, Russia*

[3]*Department of Electrical and Computer Engineering, Colorado State University,*
*Fort Collins, CO80523, USA*

[4]*Department of Physics, Colorado State University,*
*Fort Collins, CO80523, USA*

[5]*Department of Physics and Astronomy, Texas A&M University,*
*College Station, 578 University Drive, Texas 77843-4242, USA*

In [I.R. Khairulin et al., submitted to Phys. Rev. Lett.] we propose a method for amplifying a train of sub-femtosecond pulses of circularly or elliptically polarized extreme ultraviolet (XUV) radiation, constituted by high-order harmonics of an infrared (IR) laser field, in a neon-like active medium of a plasma-based X-ray laser, additionally irradiated with a replica of a fundamental frequency laser field used to generate harmonics, and show the possibility of maintaining or enhancing the ellipticity of high-harmonic radiation during its amplification. In the present paper we describe this process in detail both for a single harmonic component and a sub-femtosecond pulse train formed by a set of harmonics. We derive the analytical theory and describe both analytically and numerically the evolution of the high-harmonic field during its propagation through the medium. We discuss also the possibility of an experimental implementation of the suggested technique in an active medium of an X-ray laser based on neon-like $Ti^{12+}$ ions irradiated by an IR laser field with a wavelength of 3.9 μm.

## I. INTRODUCTION

Sub-femtosecond pulses of extreme ultraviolet (XUV) and X-ray radiation produced via high-harmonic generation (HHG) of optical laser fields have enabled for the first time an experimental study of ultrafast electronic processes in atoms, molecules, and solids on their intrinsic time scales [1-4]. However, typically such pulses are linearly polarized, which prevents from using them for the investigation of chiral and magnetic media. Generation of elliptically and circularly polarized high-order harmonics (HHs) of optical radiation is a difficult task for a number of reasons.

First, the harmonic generation efficiency decreases rapidly with increasing ellipticity of the fundamental laser field [5, 6], because, in accordance with the semiclassical model of HHG [7], when the polarization of the electric field differs from linear, the freed electrons move along curved paths; in this case, the electron trajectories, which at the moment of release from the atom had a transverse velocity close to zero, subsequently miss the parent ion. Moreover, the ellipticity of the generated harmonics rapidly decreases with their order. This is because the main contribution to the harmonics of an increasingly higher order is made by ever longer electron paths; in terms of quantum mechanics, such trajectories correspond to electron wave packets that experience, at the moment of return to the parent ion, an increasingly strong spreading and, accordingly, have lower transverse inhomogeneity, which entails more and more significant decrease in the induced transverse dipole moment and, therefore, ellipticity of the generated harmonics [8]. All of this prevents the use of elliptically polarized laser fields to efficiently generate harmonics with high ellipticity. The exceptions are the cases of harmonic frequencies satisfying the conditions of a resonance below the ionization threshold or continuum resonance of the target. For both of these cases, the resonant HHG enhancement has been observed in [9], where the interactions of elliptically polarized laser field with Ar atoms and $SF_6$ molecules resulted in enhanced elliptically polarized harmonic production due to, respectively, resonances below and above the ionization threshold $I_p$. The maximum ellipticity of resonantly enhanced harmonics in both cases reached 0.75–0.8. It should be noted, however, that such a resonant enhancement was observed only in a narrow interval of harmonic photon energies near 15.3 eV (Ar) and in the range of 20-27 eV ($SF_6$).

One of the alternative approaches to producing elliptically or circularly polarized HHs involves the generation of linearly polarized harmonics upon irradiation of gases by linearly polarized laser field, with further conversion of the generated radiation to elliptically polarized using phase-shifting optics. For example, the use of phase retarders based on a sequence of mirrors [10] and multilayer quarter waveplates [11] was demonstrated experimentally. Among the drawbacks of the demonstrated schemes of this kind are their low (about few percent) transmission efficiency and narrow frequency tuning region (not exceeding several eV).

A number of experimental studies have demonstrated methods for producing an elliptically polarized XUV radiation directly in the HHG process. One of the proposed approaches consists in generating elliptically polarized harmonics in a gas of aligned molecules. In [12], a linearly polarized driver was used, which ensured high efficiency of HHG; the appearance of the transverse component of the induced dipole moment in this case is due to the breakdown of the cylindrical symmetry of the system relative to the polarization axis of the laser field. However, in the experiments performed, the harmonic ellipticity thus obtained did not exceed 0.4. An experiment [13] also demonstrated the generation of elliptically polarized HHs upon irradiation of an ensemble of aligned $CO_2$ molecules by a circularly polarized laser field. Such generation is allowed due to the anisotropy of the medium introduced by the aligning pulse. However, high ellipticity was achieved in [13] only for several low-order harmonics; furthermore, the harmonic yield in this case is at least two orders of magnitude lower than in the usual case of linearly polarized driving field, which limits the application of this approach to high peak power lasers.

Much attention is paid to HHG schemes using multicomponent laser fields, whose components have different carrier frequencies and intensities, as well as, in the general case, polarizations. For example, recent experiments [14, 15] demonstrated the generation of harmonics with ellipticity reaching 0.7-0.75 in noble gases exposed to a two-component laser field, which is a superposition of fundamental-frequency and second-harmonic fields with mutually orthogonal linear polarizations. It turns out, however, that in the case of low-intensity second-harmonic field [14], high ellipticity is achieved only for even-order harmonics, whose intensity is low. In the case when the intensity of the second-harmonic field was scanned [15], the even- and odd-order harmonics of comparable intensity were obtained. However, it turns out that in this scheme the harmonics radiated from two adjacent half optical cycles have opposite helicity; moreover, the even- and odd-order harmonics have different polarization directions. All this makes the above schemes unfavorable for the synthesis of bright attosecond pulses with a well-defined polarization state.

In recent years, of great interest has been the use of the so-called bielliptic or bicircular fields, in which the polarization of both co-propagating beams, whose frequencies are usually related as 1:2 (ω-2ω scheme), are, respectively, elliptical [16] or circular [17], whereas the electric field vector rotates in opposite directions. This field combination, according to the energy- and spin angular momentum conservation, gives rise to circularly polarized harmonics, where, in the ω-2ω scheme, the harmonics of orders q=3m−1 rotate right and the harmonics of orders q=3m+1 rotate left, whereas the 3m harmonics are, theoretically, completely suppressed. As a drawback of this scheme, it should be noted that the strong alternation of helicities between adjacent harmonics makes it difficult to synthesize ultrashort circularly polarized pulses. Furthermore, the harmonics emitted by this source, strictly speaking, are not exactly circularly polarized, but only on average. In fact, they are trains of linearly polarized attosecond pulses with alternating direction of polarization, and therefore cannot be directly used to probe the processes in chiral media on an attosecond time scale.

Despite the shortcomings described above, a number of experiments have already demonstrated the use of these sources for diagnostics of magnetic and chiral media [9, 14, 17, 18], which allows one to declare the emergence of a new type of instrumentation. However, for all the reasons stated above, an important task is to amplify elliptically and circularly polarized high-harmonic radiation, especially in the so far poorly accessible region of photon energies of about 100 eV or more. Recently, such a possibility was experimentally demonstrated for the case of the active medium of a plasma X-ray laser based on nickel-like $Kr^{8+}$ ions [19]. At the same time, in this experiment, only one resonant harmonic with a wavelength of 32.8 nm was amplified, and the ellipticity of the amplified radiation was completely determined by the ellipticity of the seed radiation.

Our recent work [20] has shown the possibility to amplify attosecond pulses formed by a combination of linearly polarized HHs of an IR laser field in a hydrogen-like active medium of a plasma-based X-ray laser simultaneously irradiated with a replica of a fundamental-frequency IR field with the same linear polarization. Under the action of an intense IR field, due to the linear Stark effect caused by the degeneracy of the $|2s\rangle$ and $|2p\rangle$ states in the hydrogen-like ions, the gain spectrum of the active medium is redistributed from the frequency of the inverted transition to the sidebands whose frequencies are distant from the resonance by an even multiple of the modulation frequency and hence coincide with the frequencies of HHs forming the attosecond pulses. However, because only two out of four energy degenerate states, corresponding to an excited energy level, experience linear Stark effect, in the hydrogen-like medium the gain redistribution occurs only for the XUV/X-ray field with the same linear polarization as that of the modulating field (we choose the $x$-coordinate axis to be parallel to the propagation direction of both the XUV and IR fields, and $z$-axis to be parallel to the polarization direction of the IR field). For the XUV radiation with the orthogonal polarization (along the $y$-axis) the gain remains localized at the (single) resonance frequency, which prevents the amplification of elliptically or circularly polarized harmonics.

In [21], we show that amplification of sub-femtosecond pulses of the radiation of a set of harmonics with an arbitrary elliptical (including circular) polarization is possible in a neon-like active medium of a plasma X-ray laser modulated by a linearly polarized IR laser field. In this case, the major effect of the IR laser field is a fast (in a fraction of the optical cycle) shift of the energy levels of the resonant ions due to the quadratic Stark effect. As a result, unlike the case of a hydrogen-like active medium, the gain redistribution to combination frequencies occurs for the XUV/X-ray radiation with linear polarization both parallel and orthogonal to the polarization direction of the IR field. At a certain intensity of the modulating field, the frequencies of the induced gain lines coincide for the $z$- and $y$-polarization components of the XUV/X-ray radiation, which makes it possible to amplify the set of circularly or elliptically polarized harmonics. In this case, two characteristic regions are formed in the gain spectrum of the active medium. In one of them, the gain for different polarization components of the XUV/X-ray radiation is almost the same, which makes it possible to amplify sub-femtosecond pulse trains of HHs without significantly changing the shape, duration, and polarization of the pulses. At the same time, in the other part of the spectrum, the gain for one of the polarization components is much higher than for orthogonally polarized radiation, which allows one to control the harmonic ellipticity (and, in particular, to increase it).

In this work, we address the amplification process of elliptically/circularly polarized HHs in more detail. In particular, we derive the analytical theory and, based on both the analytical solution and numerical studies, discuss the evolution of the high-harmonic field during its propagation through the medium. The possibilities of the experimental implementation of this approach are dis-cussed here on the example of an active medium of nickel-like Ti$^{12+}$ ions. The generation of radiation with a wavelength of 32.6 nm in a plasma of inverted Ti$^{12+}$ ions was actively studied in [22-25]; see also review [26]. In addition, [27] showed the possibility of using neon-like Ti$^{12+}$ ions to amplify the resonant high-order harmonic radiation with linear polarization. In a number of studies, the generation of XUV/X-rays in the plasma of other neon-like ions with a similar energy level structure was also studied, see [28-32] and reviews [26, 33, 34]. In addition, the fundamental scheme of energy levels coincides for many neon- and nickel-like ions, in particular, for neon-like Ti$^{12+}$ ions and nickel-like Kr$^{8+}$ ions (used in [19] to amplify the resonant harmonic of circular polarization) or nickel-like Mo$^{14+}$ and Ag$^{19+}$ ions [35]. Thus, potentially the results of this work could be applied to amplification of harmonics of elliptical or circular polarization at different wavelengths using active media of plasma x-ray lasers based on various neon-like or nickel-like ions.

This paper is organized as follows. The theoretical model is described in Sec. II. Section III presents an analytical solution for a single high-order harmonic in a modulated neon-like active medium of a plasma-based X-ray laser, on the basis of which the optimal conditions are discussed for amplification of a single harmonic (a) with and (b) without significant change of polarization. Further, in Sec. IV, the results of numerical calculations are presented, which confirm the possibility of amplification and controlling the polarization state of a single harmonic or a combination of HHs with elliptical or circular polarization. Finally, in Sec. V, the conclusions are given.

## II. THEORETICAL MODEL

Below we consider the amplification of the harmonics of the optical field in the active medium of a plasma-based X-ray laser with inversion at the 3p$^1$S$_0$- 3s$^1$P$_1$ transition of neon-like Ti$^{12+}$ ions with the unperturbed resonance wavelength of 32.6 nm [22-27]. The scheme of energy levels of the 3p$^1$S$_0$- 3s$^1$P$_1$ transition of Ti$^{12+}$ions is shown in Fig. 1. The upper level is nondegenerate and corresponds to the state $|1\rangle$ with the total momentum value $J = 0$. The lower energy level is triply degenerate and corresponds to the $|2\rangle$, $|3\rangle$, and $|4\rangle$ states with $J = 1$ and the momentum projection on the quantization axis $M = 0$, $M = 1$, and $M = -1$, respectively. Further, we assume that the active medium has the form of a thin cylinder oriented along the $x$ axis, and use the spatially one-dimensional approximation implying that the characteristics of the medium and fields change only along the $x$ axis. We choose the $z$ axis as the quantization axis. In this case, the dipole moment of the transition from the state $|1\rangle$ to the state $|2\rangle$ will be oriented along the $z$ axis, while the dipole moments of the transitions between the states $|1\rangle$ and $|3\rangle$, as well as between the states $|1\rangle$ and $|4\rangle$, will have components along the $x$ and $y$ axes:

$$\vec{d}_{21} = \vec{z}_0 d_z, \ \vec{d}_{31} = -\vec{x}_0 d_x - i\vec{y}_0 d_y, \ \vec{d}_{41} = \vec{x}_0 d_x - i\vec{y}_0 d_y, \tag{1}$$

where $\vec{x}_0$, $\vec{y}_0$, and $\vec{z}_0$ are unit vectors along the *x*, *y*, and *z* axes, respectively; $d_z = D/\sqrt{3}$ and $d_x = d_y = D/\sqrt{6}$, whereas $D \equiv \left|\langle 3p^1S_0 \| D \| 3s^1P_1 \rangle\right|$ is the reduced matrix element of the transition dipole moment, see [36], which, according to [37], can be calculated through the spontaneous emission rate at the inverted transition, $W(3p^1S_0; 3s^1P_1)$, as $D = \sqrt{\dfrac{3\hbar c^3}{4\omega^3} W(3p^1S_0; 3s^1P_1)}$.

The active medium of a plasma-based X-ray laser is irradiated with a resonant radiation from the seed of the XUV range, which at the input to the medium, $x = 0$, has the form

$$\vec{E}^{(inc)}(t) = \frac{1}{2}\left(\vec{z}_0 \tilde{E}_z^{(inc)}(t) + \vec{y}_0 \tilde{E}_y^{(inc)}(t)\right)\exp\{-i\omega t\} + \text{c.c.}, \qquad (2)$$

where $\omega$ is the carrier frequency of the XUV radiation, $\tilde{E}_z^{(inc)}(t)$ and $\tilde{E}_y^{(inc)}(t)$ are slowly varying complex amplitudes of the components of the incident field polarized along the *z* and *y* axes, respectively, and c.c. denotes a complex conjugate. We will further consider the amplification of quasimonochromatic resonant radiation (in Sec. III), as well as the amplification of the set of harmonics of the IR field that form a sequence of short pulses (in Sec. IV). We also note that in Eq. (2) the phase difference between complex functions $\tilde{E}_z^{(inc)}(t)$ and $\tilde{E}_y^{(inc)}(t)$ can be arbitrary, which allows one to consider XUV radiation with an arbitrary (in the general case, elliptical) polarization.

In addition to XUV radiation, the active medium of a plasma-based X-ray laser is irradiated with an intense IR laser field propagating along the *x* axis and polarized along the *z* axis:

$$\vec{E}_{IR}(x,t) = \vec{z}_0 E_M \cos\left(\Omega\left[t - \frac{\sqrt{\varepsilon_{pl}^{(IR)}}}{c}x\right]\right). \qquad (3)$$

Here $E_M$ and $\Omega$ are the amplitude and frequency of the IR field, *c* is the speed of light in vacuum, $\varepsilon_{pl}^{(IR)} = 1 - \omega_{pl}^2/\Omega^2$ is the dielectric constant of the plasma for the modulating field, $\omega_{pl} = \sqrt{4\pi N_e e^2/m_e}$ is the electron plasma frequency (hereinafter, the formulas are given in the CGS-ESU system), $N_e$ is the concentration of free electrons in the plasma, and *e* and $m_e$ are the charge and mass of the electron, respectively. In Eq. (3), the pulse duration of the IR field is assumed to be significantly longer than the duration of all the processes under study, which allows us to consider it monochromatic. In the case of Ti$^{12+}$ ions, the frequency of a mid-IR laser field is at least two orders of magnitude lower than the frequency of any of the transitions connecting all four states |1⟩-|4⟩, shown in Fig.1, with other ionic quantum states, as well as the transition frequencies from the upper state |1⟩ to the lower states |2⟩, |3⟩, and |4⟩ (while the transitions between those lower states are forbidden). As a result, the influence of the IR field on the |1⟩-|4⟩ states in such adiabatic perturbation

regime is reduced predominantly to a shift of the corresponding energy levels due to the quadratic Stark effect. In this case, the position of the $i$-th energy level ($i = 1,2,3,4$) is determined by [38]

$$E_i(x,t) = E_i^{(0)} + \frac{1}{2}\sum_{k \neq i} \frac{|d_{ki}^{(z)}|^2 E_M^2}{\hbar \omega_{ik}} \left(1 + \cos\left[2\Omega\left(t - \frac{\sqrt{\varepsilon_{pl}^{(IR)}}}{c} x\right)\right]\right), \quad (4)$$

where $E_i^{(0)}$ is the unperturbed energy value, $\omega_{ik}$ is the unperturbed frequency of the transition from the $i$ state to the $k$ state, and $d_{ki}^{(z)}$ is the projection of the dipole moment of a given transition on the $z$ axis. The summation in (4) is carried out over several dozens of states of the unperturbed (field-free) ion with the largest electric dipole moments of transitions to the state $|i\rangle$ (including the states $|1\rangle$-$|4\rangle$) [39]. We further introduce the notations $\Delta_E^{(i)} = \frac{1}{2}\sum_{k \neq i} \frac{|d_{ki}^{(z)}|^2 E_M^2}{\hbar \omega_{ik}}$ for the amplitude of the shift of the energy level corresponding to the $i$-th state and $\Delta_\Omega^{(ij)} = \left(\Delta_E^{(i)} - \Delta_E^{(j)}\right)/\hbar$ for the amplitude of the frequency change of the $|i\rangle \leftrightarrow |j\rangle$ transition, where $i, j = 1,2,3,4$. With the notations introduced, the instantaneous values of the frequencies of quantum transitions between the $|1\rangle$, $|2\rangle$, $|3\rangle$, and $|4\rangle$ states can be represented as

$$\begin{cases} \omega_{12}(\tau, x) = \bar{\omega}_{tr}^{(z)} + \Delta_\Omega^{(z)} \cos[2\{\Omega\tau + \Delta K x\}], \\ \omega_{13}(\tau, x) = \bar{\omega}_{tr}^{(y)} + \Delta_\Omega^{(y)} \cos[2\{\Omega\tau + \Delta K x\}], \\ \omega_{14}(\tau, x) = \omega_{13}(\tau, x), \\ \omega_{23}(\tau, x) = \left(\Delta_\Omega^{(z)} - \Delta_\Omega^{(y)}\right)\left(1 + \cos[2\{\Omega\tau + \Delta K x\}]\right), \\ \omega_{24}(\tau, x) = \omega_{23}(\tau, x), \\ \omega_{34}(\tau, x) = 0, \end{cases} \quad (5)$$

where $\tau = t - x\sqrt{\varepsilon_{pl}^{(XUV)}}/c$ is the local time in the reference frame moving along the $x$ axis with the phase velocity of the XUV radiation (2) in the plasma, $\varepsilon_{pl}^{(XUV)} = 1 - \omega_{pl}^2/\omega^2$ is the dielectric constant of the plasma for the XUV radiation, $\Delta K = \Omega\left(\sqrt{\varepsilon_{pl}^{(XUV)}} - \sqrt{\varepsilon_{pl}^{(IR)}}\right)/c$ is an addition to the wave number of the modulating field due to the difference between its phase velocity and the XUV radiation phase velocity, $\bar{\omega}_{tr}^{(z)} \equiv \left(E_1^{(0)} - E_2^{(0)}\right)/\hbar + \Delta_\Omega^{(z)}$ is the time-averaged $|1\rangle\leftrightarrow|2\rangle$ transition frequency, $\bar{\omega}_{tr}^{(y)} \equiv \left(E_1^{(0)} - E_3^{(0)}\right)/\hbar + \Delta_\Omega^{(y)} = \left(E_1^{(0)} - E_4^{(0)}\right)/\hbar + \Delta_\Omega^{(y)}$ is the time-averaged frequency of $|1\rangle\leftrightarrow|3\rangle$ and $|1\rangle\leftrightarrow|4\rangle$ transitions, $\Delta_\Omega^{(z)} \equiv \Delta_\Omega^{(12)}$, and $\Delta_\Omega^{(y)} \equiv \Delta_\Omega^{(13)} = \Delta_\Omega^{(14)}$.

Two important conclusions follow from Eqs. (4) and (5). First, unlike in an active medium of hydrogen-like ions [20], where the IR field linearly polarized along the $z$ axis causes the linear Stark effect and leads to spatiotemporal modulation of frequencies of only those transitions, which are involved in the interaction with the *z*-polarized XUV / X-ray radiation, in a medium of neon-like Ti$^{12+}$ ions, due to the quadratic Stark effect, the frequencies of transitions interacting with both the *z*-polarized and *y*-polarized components of the XUV radiation get modulated. Second, in the presence of a modulating IR field, both the average values and the amplitudes of the frequency shift of the $|1\rangle \leftrightarrow |2\rangle$ and $|1\rangle \leftrightarrow |3\rangle, |4\rangle$ quantum transitions interacting with *z*- and *y*-polarized components of XUV radiation are different, since the states $|2\rangle$ and $|3\rangle, |4\rangle$ experience different energy shifts due to the IR field-induced quadratic Stark effect, see Fig.1.

The evolution of the quantum state of Ti$^{12+}$ ions under the influence of XUV radiation (2) is described by the equations for the density matrix elements $\rho_{ij}$ (von Neumann equations with relaxation):

$$\begin{cases} \dfrac{\partial \rho_{11}}{\partial t} + \gamma_{11}\rho_{11} = \dfrac{i}{\hbar}\sum_{s=1}^{4}\left(\rho_{s1}\vec{d}_{1s} - \rho_{1s}\vec{d}_{s1}\right)\vec{E}, \\ \dfrac{\partial \rho_{ii}}{\partial t} + \gamma_{ii}\rho_{ii} = A\rho_{11} + \dfrac{i}{\hbar}\sum_{s=1}^{4}\left(\rho_{s1}\vec{d}_{1s} - \rho_{1s}\vec{d}_{s1}\right)\vec{E}, \text{ if } i \neq 1, \\ \dfrac{\partial \rho_{ij}}{\partial t} + \left(i\omega_{ij}(t,x) + \gamma_{ij}\right)\rho_{ij} = \dfrac{i}{\hbar}\sum_{s=1}^{4}\left(\rho_{sj}\vec{d}_{is} - \rho_{is}\vec{d}_{sj}\right)\vec{E}, \text{ if } j \neq i, \end{cases} \quad (6)$$

where $\vec{E} = \vec{E}(x,t)$ is the XUV radiation in the medium, $\omega_{ij}(t,x)$ are the frequencies of quantum transitions determined by Eqs. (5), $A$ is the rate of spontaneous radiative transition from the $|1\rangle$ state to each of the $|2\rangle, |3\rangle$, and $|4\rangle$ states, $1/A = 242.5$ ps [39], and $\gamma_{ij}$ are the relaxation rates of the density-matrix elements, which are determined as follows. The relaxation rate of the density matrix diagonal element $\rho_{ii}$ is the sum of the rates of radiative transitions from the $|i\rangle$ state to all lower-energy ionic states, $\Gamma_{rad}^{(i)}$, and the rate of tunneling ionization from the $|i\rangle$ state under the influence of the IR field, $w_{ion}^{(i)}$ (which is calculated using the Perelomov-Popov-Terent'ev formula [40]), i.e. $\gamma_{ii} = \Gamma_{rad}^{(i)} + w_{ion}^{(i)}$. Note that in the range of IR intensities under consideration, ionization from resonant states is not significant, $w_{ion}^{(i)} < \Gamma_{rad}^{(i)}$ $(\forall i)$. The relaxation rate of an off-diagonal element $\rho_{ij}$ of the density matrix is determined as $\gamma_{ij} = (\gamma_{ii} + \gamma_{jj})/2 + \gamma_{Coll}$, where $\gamma_{Coll}$ is the frequency of collisions in the plasma.

Next, we turn to local time $t \to \tau = t - x\sqrt{\varepsilon_{pl}^{(XUV)}}/c$ and look for a solution of the system of equations (6) using the approximation of slowly varying amplitudes, i.e. assuming that

$$\vec{E}(x,\tau) = \frac{1}{2}\left(\vec{z}_0\tilde{E}_z(x,\tau) + \vec{y}_0\tilde{E}_y(x,\tau)\right)\exp\{-i\omega\tau\} + \text{c.c.}, \quad (7)$$
$$\rho_{12}(x,\tau) = \tilde{\rho}_{12}(x,\tau)e^{-i\omega\tau}, \quad \rho_{13}(x,\tau) = \tilde{\rho}_{13}(x,\tau)e^{-i\omega\tau}, \quad \rho_{14}(x,\tau) = \tilde{\rho}_{14}(x,\tau)e^{-i\omega\tau},$$

$$\rho_{ij} = \rho_{ji}^* \text{ and } \rho_{ij}(x,\tau) = \tilde{\rho}_{ij}(x,\tau) \text{ if } ij \neq \{12, 21, 13, 31, 14, 41\},$$

where $\tilde{E}_z(x,\tau)$ and $\tilde{E}_y(x,\tau)$ are slowly varying complex amplitudes of the polarization components of the XUV radiation in the medium, $\left|\frac{\partial \tilde{E}_{z,y}}{\partial \tau}\right| \ll \omega |\tilde{E}_{z,y}|$ and $\left|\frac{\partial \tilde{E}_{z,y}}{\partial x}\right| \ll \frac{\omega \sqrt{\varepsilon_{pl}^{(XUV)}}}{c} |\tilde{E}_{z,y}|$, whereas $\tilde{\rho}_{ij}(x,\tau)$ are slowly varying amplitudes of the density-matrix elements of the medium, $\left|\frac{\partial \tilde{\rho}_{ij}}{\partial \tau}\right| \ll \omega |\tilde{\rho}_{ij}|$ and $\left|\frac{\partial \tilde{\rho}_{ij}}{\partial x}\right| \ll \frac{\omega \sqrt{\varepsilon_{pl}^{(XUV)}}}{c} |\tilde{\rho}_{ij}|$. Assuming that the radiation frequency of (2) is close to the transition frequencies $|1\rangle \leftrightarrow |2\rangle$ and $|1\rangle \leftrightarrow |3\rangle, |4\rangle$, which displacement under the influence of the IR field is taken into account, and using the rotating wave approximation, the system of Eqs. (6) can be represented as

$$\begin{cases}
\frac{\partial \tilde{\rho}_{11}}{\partial \tau} + \gamma_{11}\tilde{\rho}_{11} = \frac{i}{2\hbar}d_z\left(\tilde{\rho}_{12}^*\tilde{E}_z - \tilde{\rho}_{12}\tilde{E}_z^*\right) - \frac{1}{2\hbar}d_y\left(\tilde{\rho}_{13}^*\tilde{E}_y + \tilde{\rho}_{13}\tilde{E}_y^*\right) - \frac{1}{2\hbar}d_y\left(\tilde{\rho}_{14}^*\tilde{E}_y + \tilde{\rho}_{14}\tilde{E}_y^*\right), \\
\frac{\partial \tilde{\rho}_{22}}{\partial \tau} + \gamma_{22}\tilde{\rho}_{22} = A\tilde{\rho}_{11} - \frac{i}{2\hbar}d_z\left(\tilde{\rho}_{12}^*\tilde{E}_z - \tilde{\rho}_{12}\tilde{E}_z^*\right), \\
\frac{\partial \tilde{\rho}_{33}}{\partial \tau} + \gamma_{33}\tilde{\rho}_{33} = A\tilde{\rho}_{11} + \frac{1}{2\hbar}d_y\left(\tilde{\rho}_{13}^*\tilde{E}_y + \tilde{\rho}_{13}\tilde{E}_y^*\right), \\
\frac{\partial \tilde{\rho}_{44}}{\partial \tau} + \gamma_{44}\tilde{\rho}_{44} = A\tilde{\rho}_{11} + \frac{1}{2\hbar}d_y\left(\tilde{\rho}_{14}^*\tilde{E}_y + \tilde{\rho}_{14}\tilde{E}_y^*\right), \\
\frac{\partial \tilde{\rho}_{12}}{\partial \tau} + \left[i(\omega_{12}(\tau,x) - \omega) + \gamma_{12}\right]\tilde{\rho}_{12} = -\frac{i}{2\hbar}d_z(\tilde{\rho}_{11} - \tilde{\rho}_{22})\tilde{E}_z - \frac{1}{2\hbar}d_y\tilde{\rho}_{23}^*\tilde{E}_y - \frac{1}{2\hbar}d_y\tilde{\rho}_{24}^*\tilde{E}_y, \\
\frac{\partial \tilde{\rho}_{13}}{\partial \tau} + \left[i(\omega_{13}(\tau,x) - \omega) + \gamma_{13}\right]\tilde{\rho}_{13} = \frac{1}{2\hbar}d_y(\tilde{\rho}_{11} - \tilde{\rho}_{33})\tilde{E}_y + \frac{i}{2\hbar}d_z\tilde{\rho}_{23}\tilde{E}_z - \frac{1}{2\hbar}d_y\tilde{\rho}_{34}^*\tilde{E}_y, \\
\frac{\partial \tilde{\rho}_{14}}{\partial \tau} + \left[i(\omega_{14}(\tau,x) - \omega) + \gamma_{14}\right]\tilde{\rho}_{14} = \frac{1}{2\hbar}d_y(\tilde{\rho}_{11} - \tilde{\rho}_{44})\tilde{E}_y + \frac{i}{2\hbar}d_z\tilde{\rho}_{24}\tilde{E}_z - \frac{1}{2\hbar}d_y\tilde{\rho}_{34}\tilde{E}_y, \\
\frac{\partial \tilde{\rho}_{23}}{\partial \tau} + \left[i\omega_{23}(\tau,x) + \gamma_{23}\right]\tilde{\rho}_{23} = \frac{i}{2\hbar}d_z\tilde{\rho}_{13}\tilde{E}_z^* + \frac{1}{2\hbar}d_y\tilde{\rho}_{12}^*\tilde{E}_y, \\
\frac{\partial \tilde{\rho}_{24}}{\partial \tau} + \left[i\omega_{24}(\tau,x) + \gamma_{24}\right]\tilde{\rho}_{24} = \frac{i}{2\hbar}d_z\tilde{\rho}_{14}\tilde{E}_z^* + \frac{1}{2\hbar}d_y\tilde{\rho}_{12}^*\tilde{E}_y, \\
\frac{\partial \tilde{\rho}_{34}}{\partial \tau} + \left[i\omega_{34}(\tau,x) + \gamma_{34}\right]\tilde{\rho}_{34} = \frac{1}{2\hbar}d_y\tilde{\rho}_{14}\tilde{E}_y^* + \frac{1}{2\hbar}d_y\tilde{\rho}_{13}^*\tilde{E}_y.
\end{cases} \quad (8)$$

In turn, when XUV radiation propagates in the active medium of a plasma-based X-ray laser, its slowly varying amplitude $\vec{\tilde{E}}(x,\tau) = \vec{z}_0 \tilde{E}_z(x,\tau) + \vec{y}_0 \tilde{E}_y(x,\tau)$ changes in accordance with the equation

$$\frac{\partial \vec{\tilde{E}}}{\partial x} = i2\pi \frac{\omega}{c\sqrt{\varepsilon_{pl}^{(XUV)}}} \vec{\tilde{P}}, \quad (9)$$

which is obtained from the wave equation after changing variables $x,t \rightarrow x,\tau$ within the approximation of slowly varying amplitudes [41]. In Eq. (9), $\vec{\tilde{P}}$ is the slowly varying amplitude of the resonant polarization of the medium $\vec{P}(x,\tau) = \frac{1}{2}\vec{\tilde{P}}(x,\tau)\exp\{-i\omega\tau\} + \text{c.c.}$, which is expressed through the density matrix in accordance with the equation

$$\vec{\tilde{P}} = N_{ion}\text{Tr}\left(\hat{\vec{d}}\hat{\rho}\right) = N_{ion}\left\{\rho_{12}\vec{d}_{21} + \rho_{13}\vec{d}_{31} + \rho_{14}\vec{d}_{41}\right\} + \text{c.c.}, \tag{10}$$

where $N_{ion}$ is the concentration of Ti$^{12+}$ ions in the $|1\rangle$, $|2\rangle$, $|3\rangle$, and $|4\rangle$ states at the initial instant of time, whereas $\hat{\vec{d}}$ and $\hat{\rho}$ are the operators of dipole moment and density matrix of the medium, respectively. Using the explicit form (1) of the dipole moments of the $|1\rangle \leftrightarrow |2\rangle$ and $|1\rangle \leftrightarrow |3\rangle,|4\rangle$ transitions, we can write the equations for the polarization components of the XUV field as follows:

$$\begin{cases} \dfrac{\partial \tilde{E}_z}{\partial x} = i4\pi N_{ion} d_z \dfrac{\omega}{c\sqrt{\varepsilon_{pl}^{(XUV)}}} \tilde{\rho}_{12}, \\ \dfrac{\partial \tilde{E}_y}{\partial x} = 4\pi N_{ion} d_y \dfrac{\omega}{c\sqrt{\varepsilon_{pl}^{(XUV)}}} (\tilde{\rho}_{13} + \tilde{\rho}_{14}). \end{cases} \tag{11}$$

Further, we will assume that at $\tau = t - x\sqrt{\varepsilon_{pl}^{(XUV)}}/c = 0$, among the states $|1\rangle$- $|4\rangle$, only the state $|1\rangle$ is populated: $\tilde{\rho}_{11}(x,\tau=0) = 1$ and $\tilde{\rho}_{ii}(x,\tau=0) = 0$ for $i \neq 1$. This approximation is justified by the fact that the rate of spontaneous radiative transitions from the state $|1\rangle$ to the lower-energy states is much lower than the rates of radiative transitions from the $|2\rangle$, $|3\rangle$, and $|4\rangle$ states, namely, $1/\Gamma_{rad}^{(1)} \approx 50.2$ ps, whereas $1/\Gamma_{rad}^{(2,3,4)} \approx 3.34$ ps. As a result, during the evolution of a plasma created by a sequence of pump pulses that precede the pulses of both the resonant radiation (2) and the modulating field (3) and propagate along the $x$ axis, the resonant Ti$^{12+}$ ions accumulate in the state $|1\rangle$. Further, in order to simulate the amplified spontaneous emission (ASE) of the active medium, we set random initial (at $\tau = 0$) values of coherences (off-diagonal elements of the density matrix) at the inverted $|1\rangle \leftrightarrow |2\rangle$, $|3\rangle$, $|4\rangle$ transitions (see [42, 20, 43]), whereas the initial values of the remaining coherences are equal to zero. We also note that at $\tau = 0$ the XUV radiation at the corresponding point in the medium (with the exception of the front boundary of the medium) is absent: $\tilde{E}_z(x \neq 0, \tau = 0) = 0$ and $\tilde{E}_y(x \neq 0, \tau = 0) = 0$.

Due to the absence of reflections of XUV radiation from the boundaries of the medium, the boundary conditions at the front side of the medium have the form $\tilde{E}_z(x=0,\tau) = \tilde{E}_z^{(inc)}(\tau)$ and $\tilde{E}_y(x=0,\tau) = \tilde{E}_y^{(inc)}(\tau)$. Similarly, at the back side of the medium ($x=L$), $\tilde{E}_z^{(out)}(\tau) = \tilde{E}_z(x=L,\tau)$ and $\tilde{E}_y^{(out)}(\tau) = \tilde{E}_y(x=L,\tau)$, where $\tilde{E}_z^{(out)}$ and $\tilde{E}_y^{(out)}$ are the polarization components of the slowly varying amplitude of the XUV radiation behind the medium,

$$\vec{E}^{(out)}(\tau) = \frac{1}{2}\left(\vec{z}_0 \tilde{E}_z^{(out)}(\tau) + \vec{y}_0 \tilde{E}_y^{(out)}(\tau)\right)\exp\{-i\omega\tau\} + \text{c.c.}$$

Equations (8) and (11), together with the above-mentioned initial and boundary conditions, completely characterize the temporal dynamics and spatial evolution of resonant radiation and the quantum state of the active medium. In the next section, we consider their solution in the case of a quasimonochromatic elliptically polarized seed radiation (2).

### III. AMPLIFICATION AND CONTROL OF THE ELLIPTICITY OF QUASIMONOCHROMATIC XUV SEED RADIATION: ANALYTICAL SOLUTION

In this section, we consider amplification of the quasimonochromatic XUV radiation (2) with envelope slowly changing on the scale of the modulating IR field cycle: $\left|\frac{\partial \tilde{E}_{z,y}^{(inc)}}{\partial \tau}\right| \ll \Omega \left|\tilde{E}_{z,y}^{(inc)}\right|$.

First of all, we derive an analytical solution for the polarization components of the XUV radiation in a medium, $\tilde{E}_z(x,\tau)$ and $\tilde{E}_y(x,\tau)$, in a linear approximation (see [43]), assuming that (i) during the considered time interval the population of states does not change, $\tilde{\rho}_{ii}(x,\tau) = \tilde{\rho}_{ii}(x,\tau=0)$, $i$=1,2,3,4, (ii) the coherences on the dipole-forbidden transitions between the $|2\rangle$, $|3\rangle$, and $|4\rangle$ states are identically equal to zero, and (iii) the coherences at the dipole-allowed $|1\rangle\leftrightarrow|2\rangle$, $|1\rangle\leftrightarrow|3\rangle$, and $|1\rangle\leftrightarrow|4\rangle$ transitions are equal to zero before the arrival of the XUV field, which corresponds to neglecting the spontaneous emission of the medium. The first two approximations are valid for not very large values of local time $\tau$ and not very high intensity of the seed. At the same time, the third approximation implies that the intensity of the seed is high enough to neglect the spontaneous emission of the medium. These approximations are compatible with each other as it is confirmed by the results of the numerical integration of the nonlinear system of equations (8), (11), which are presented in the next section. In the linear approximation, the amplification of the polarization components of the XUV radiation occurs independently, therefore, for simplicity, we consider them separately. To derive an analytical solution, we assume that the seed radiation (2) is turned on at the moment $\tau = 0$ and subsequently has a constant amplitude:

$$\tilde{E}_z^{(inc)}(\tau) = \theta(\tau)E_0^{(z)} \text{ and } \tilde{E}_y^{(inc)}(\tau) = \theta(\tau)E_0^{(y)}, \tag{12}$$

where $\theta(\tau)$ is the Heaviside unit step function: $\theta(\tau) = 0$ for $\tau < 0$ and $\theta(\tau) = 1$ for $\tau \geq 0$; $E_0^{(z)}$ and $E_0^{(y)}$ are complex numbers.

In the linear approximation, the $z$-component of the XUV radiation in the medium satisfies the equations

$$\begin{cases} \dfrac{\partial \tilde{E}_z}{\partial x} = i4\pi N_{ion} d_z \dfrac{\omega}{c\sqrt{\varepsilon_{pl}^{(XUV)}}} \tilde{\rho}_{12}, \\ \dfrac{\partial \tilde{\rho}_{12}}{\partial \tau} + \left\{ i\left(\bar{\omega}_{tr}^{(z)} - \omega\right) + i\Delta_\Omega^{(z)} \cos\left[2\{\Omega\tau + \Delta K x\}\right] + \gamma_z \right\} \tilde{\rho}_{12} = -in_{tr}^{(z)} \dfrac{d_z \tilde{E}_z}{2\hbar}, \end{cases} \quad (13)$$

where $\gamma_z \equiv \gamma_{12}$ and $n_{tr}^{(z)} = \tilde{\rho}_{11}(x,\tau) - \tilde{\rho}_{22}(x,\tau)$ is the population difference at the transition $|1\rangle \leftrightarrow |2\rangle$; in the approximation under consideration, $n_{tr}^{(z)} = 1$. It is easy to show that the $z$-polarized XUV radiation will be most efficiently amplified if its frequency coincides with the time-average frequency of the $|1\rangle \leftrightarrow |2\rangle$ transition or is tuned from it to an even multiple of the modulation frequency: $\omega = \bar{\omega}_{tr}^{(z)} + 2k\Omega$, where $k$ is an integer. In this case, the solution of Eqs. (13) takes the form

$$\tilde{E}_z = E_0^{(z)} \theta(\tau) \exp\left\{ g_k^{(z)}\left(P_\Omega^{(z)}, \tau\right) x \right\}, \quad (14a)$$

$$g_k^{(z)}\left(P_\Omega^{(z)}, \tau\right) = g_{total}^{(z)} J_k^2\left(P_\Omega^{(z)}\right)\left(1 - e^{-\gamma_z \tau}\right), \quad (14b)$$

where $g_k^{(z)}\left(P_\Omega^{(z)}, \tau\right)$ is the effective gain of $z$-polarized radiation with a frequency $\omega = \bar{\omega}_{tr}^{(z)} + 2k\Omega$, $g_{total}^{(z)} = \dfrac{2\pi \omega N_{ion} n_{tr}^{(z)} d_z^2}{\hbar c \gamma_z \sqrt{\varepsilon_{pl}^{(XUV)}}}$ is the gain of $z$-polarized radiation in the absence of modulation (the gain in intensity is two times higher, $G_{total}^{(z)} = 2 g_{total}^{(z)}$), $J_k(x)$ is the Bessel function of the first kind of order $k$, and $P_\Omega^{(z)} = \dfrac{\Delta_\Omega^{(z)}}{2\Omega}$ is the frequency modulation index of the $|1\rangle \leftrightarrow |2\rangle$ transition under the influence of the IR field (due to the quadratic Stark effect). In addition to the above-mentioned approximations, the solution (14) implies an inertialess relationship between the resonant polarization of the medium and the XUV radiation, which takes place if the amplitude of the XUV radiation changes insignificantly during the time $1/\gamma_z$ until the resonant polarization of the medium settles down. This approximation is rather rough, but, as shown in [43], it qualitatively reproduces the results of numerical calculations using more general equations. In addition, the solution (14) implies that the plasma has a strong dispersion at the frequency of the IR field, $g(P_\Omega, \tau)/\Delta K \ll 1$, as a result of which the amplification of the radiation (2) is not accompanied by the widening of its spectrum (see [20]). The derivation of a similar analytical solution is described in more detail in [43].

We will further consider the amplification of the $y$-component of radiation (2). In the linear approximation, the $y$-component of the XUV radiation in the medium satisfies the equations

$$\begin{cases} \dfrac{\partial \tilde{E}_y}{\partial x} = 4\pi N_{ion} d_y \dfrac{\omega}{c\sqrt{\varepsilon_{pl}^{(XUV)}}} \left( \tilde{\rho}_{13} + \tilde{\rho}_{14} \right), \\[2mm] \dfrac{\partial \tilde{\rho}_{13}}{\partial \tau} + \left\{ i\left( \bar{\omega}_{tr}^{(y)} - \omega \right) + i\Delta_{\Omega}^{(y)} \cos\left[ 2\{\Omega\tau + \Delta Kx\} \right] + \gamma_y \right\} \tilde{\rho}_{13} = n_{tr}^{(y)} \dfrac{d_y \tilde{E}_y}{2\hbar}, \\[2mm] \dfrac{\partial \tilde{\rho}_{14}}{\partial \tau} + \left\{ i\left( \bar{\omega}_{tr}^{(y)} - \omega \right) + i\Delta_{\Omega}^{(y)} \cos\left[ 2\{\Omega\tau + \Delta Kx\} \right] + \gamma_y \right\} \tilde{\rho}_{14} = n_{tr}^{(y)} \dfrac{d_y \tilde{E}_y}{2\hbar}, \end{cases} \quad (15)$$

where $\gamma_z \equiv \gamma_{13} = \gamma_{14}$ and $n_{tr}^{(y)} = \tilde{\rho}_{11}(x,\tau) - \tilde{\rho}_{33}(x,\tau) = \tilde{\rho}_{11}(x,\tau) - \tilde{\rho}_{44}(x,\tau)$ is the population difference at the $|1\rangle \leftrightarrow |3\rangle$ and $|1\rangle \leftrightarrow |4\rangle$ transitions, which in the case under consideration is the same and equal to $n_{tr}^{(y)} = 1$. Similarly to $z$-polarized radiation, the $y$-polarized component of the XUV field is amplified most efficiently if the carrier frequency of the field coincides with the $|1\rangle \leftrightarrow |3\rangle, |4\rangle$ transition frequency or is tuned from it by an even multiple of the modulation frequency: $\omega = \bar{\omega}_{tr}^{(y)} + 2k'\Omega$, where $k'$ is an integer. In this case, the solution of Eqs. (15) has the form (14) up to a replacement of the indices $z \to y$ and $k \to k'$:

$$\tilde{E}_y = E_0^{(y)} \theta(\tau) \exp\left\{ g_y\left( P_{\Omega}^{(y)}, \tau \right) x \right\}, \quad (16a)$$

$$\text{where } g_{k'}^{(y)}\left( P_{\Omega}^{(y)}, \tau \right) = g_0^{(y)} J_{k'}^2\left( P_{\Omega}^{(y)} \right)\left( 1 - e^{-\gamma_y \tau} \right). \quad (16b)$$

Here $g_{k'}^{(y)}\left( P_{\Omega}^{(y)}, \tau \right)$ is the effective gain of the $y$-polarized radiation with frequency $\omega = \bar{\omega}_{tr}^{(y)} + 2k'\Omega$, $g_{total}^{(y)} = \dfrac{4\pi\omega N_{ion} n_{tr}^{(y)} d_y^2}{\hbar c \gamma_y \sqrt{\varepsilon_{pl}^{(XUV)}}}$ is the gain (in the field amplitude) in the absence of modulation (the gain in intensity is equal to $G_{total}^{(y)} = 2g_{total}^{(y)}$), and $P_{\Omega}^{(y)} = \dfrac{\Delta_{\Omega}^{(y)}}{2\Omega}$ is the index of modulation of the $|1\rangle \leftrightarrow |3\rangle, |4\rangle$ transition frequencies.

Due to the fact that under the influence of a modulating field, the frequencies of $|1\rangle \leftrightarrow |2\rangle$ and $|1\rangle \leftrightarrow |3\rangle, |4\rangle$ quantum transitions become different, in the general case, the conditions for resonant amplification of $z$- and $y$-polarized radiation do not coincide. However, in order for the elliptically or circularly polarized radiation to be amplified, both of its polarization components must be amplified simultaneously. This is possible if $\omega = \bar{\omega}_{tr}^{(z)} + 2k\Omega = \bar{\omega}_{tr}^{(y)} + 2k'\Omega$, i.e. the difference between the frequencies $\bar{\omega}_{tr}^{(z)}$ and $\bar{\omega}_{tr}^{(y)}$ must be equal to or a multiple of twice the frequency of the modulating field: $\bar{\omega}_{tr}^{(y)} - \bar{\omega}_{tr}^{(z)} = 2\Omega(k - k')$. We further consider the case that is easiest for experimental implementation:

$$\bar{\omega}_{tr}^{(y)} - \bar{\omega}_{tr}^{(z)} = 2\Omega. \quad (17)$$

In this case, $k' = k-1$, and if the effective gain of the z-component of the radiation is proportional to $J_k^2\left(P_\Omega^{(z)}\right)$, then the gain of the y-component of the radiation is proportional to $J_{k-1}^2\left(P_\Omega^{(y)}\right)$. Due to the fact that $\bar{\omega}_{tr}^{(y)} - \bar{\omega}_{tr}^{(z)} = \Delta_\Omega^{(y)} - \Delta_\Omega^{(z)}$ and $\Delta_\Omega^{(y)} - \Delta_\Omega^{(z)} \sim E_M^2$, see Eq. (4), the equality (17) unambiguously connects the intensity of the modulating field with its frequency and wavelength. For $Ti^{12+}$ ions, this dependence is shown in Fig. 2, where, in addition to the wavelength of the modulating field, for which the condition (17) is satisfied (left vertical axis), the ionization rates from the $|1\rangle$-$|4\rangle$ states, $w_{ion}^{(i)}$, are shown versus the intensity of the modulating field (right vertical axis). The ionization rates are normalized to the half-width of the gain line in the absence of a modulating field, $\gamma_{21}^{(0)}$. In order for the resonant polarization of the medium to have a time to develop, it is necessary to fulfill the condition $w_{ion}^{(i)} \leq \gamma_{21}^{(0)}$ that limits the maximum intensity of the modulating field and its minimum wavelength to $1.2\times10^{17}$ W/cm$^2$ and 2.7 μm, respectively. It is worth to note that the ionization rates were calculated using the Perelomov-Popov-Terent'ev formula [40], which has limited accuracy for states of multiply charged ions. Thus, the calculated values of maximum intensity and minimum wavelength of the modulating field should be considered as an approximate estimate. We further consider the modulation of the active medium by an IR field with a wavelength of 3.9 μm. The required intensity of the modulating field is $8.26\times10^{16}$ W/cm$^2$. Note that this combination of the IR field wavelength and intensity is experimentally accessible [44]. In this case, the ionization rate from the $|1\rangle$ state is equal to $1/w_{ion}^{(1)} \approx 124$ ps, and the time-averaged $|1\rangle \leftrightarrow |2\rangle$ transition frequency coincides with the frequency of 155-th harmonic of the modulating field, $\bar{\omega}_{tr}^{(z)} = 155\times\Omega$. The wavelength of resonant radiation in vacuum in this case is equal to $\lambda_{H155} = 25.17$ nm. Note that this is a calculated value obtained from the available theoretical data on the positions of the energy levels of $Ti^{12+}$ ions and the rates of radiation transitions between them [39]. According to these data, in the absence of a modulating field, the wavelength of the resonant XUV radiation is 30.04 nm, whereas in the experiment, the central wavelength of the generated radiation is 32.63 nm. This difference should be kept in mind when planning an experiment. As follows from the solutions given by Eqs. (14) and (16), the effective gain coefficients for the z- and y-polarized components of the XUV field are proportional to the squares of the Bessel functions, the order of which is determined by the detuning of the field frequency from the central frequencies of the transitions $\bar{\omega}_{tr}^{(z)}$ and $\bar{\omega}_{tr}^{(y)}$, respectively. We also note that in the linear amplification regime, under the condition $n_{tr}^{(y)} = n_{tr}^{(z)}$, the equality $g_{total}^{(y)} \simeq g_{total}^{(z)} \equiv g_{total}$ holds due to the facts that $\gamma_y \simeq \gamma_z \equiv \gamma_{tr}$ (up to a difference in the ionization rates from the $|2\rangle$ and $|3\rangle$, $|4\rangle$ states) and $2d_y^2 = d_z^2$. As a result, the difference between the effective gains for the y- and z-polarized components of the XUV radiation is solely due to the difference in the orders and arguments of the Bessel functions in Eqs. (14b) and (16b). For the indicated intensity and wavelength of the IR field, the modulation indices for the $|1\rangle \leftrightarrow |2\rangle$ and $|1\rangle \leftrightarrow |3\rangle, |4\rangle$ transitions are $P_\Omega^{(z)} \approx 12.57$ and $P_\Omega^{(y)} \approx 13.57$.

The corresponding gain spectra for the $z$- and $y$-polarized components of the XUV radiation (dependences of the effective gains $g_k^{(z)}\left(P_\Omega^{(z)}, \tau\right)$ and $g_{k-1}^{(y)}\left(P_\Omega^{(y)}, \tau\right)$ on $k = \dfrac{\omega - \bar{\omega}_{tr}^{(z)}}{2\Omega}$) are shown in Fig. 3. Note that Fig. 3 shows the values of the effective gains at $\tau \gg 1/\gamma_y, 1/\gamma_z$. In accordance with Eqs. (14b) and (16b), for smaller values of local time, the dependences of the effective gain coefficients on $k$ remain the same, but the gain amplitudes become smaller. Due to the fact that, at the same field frequency $\omega$, the $z$-component of the field is in resonance with the $k$-th induced gain line of the active medium, and the $y$-component of the field is in resonance with the ($k$ - 1)-th gain line, and also due to the differences in the modulation indices for the $|1\rangle \leftrightarrow |2\rangle$ and $|1\rangle \leftrightarrow |3\rangle, |4\rangle$ transitions, the gain coefficients for the $z$- and $y$-components of the XUV radiation are different. In this case, depending on the value of $k$, any of the three options is possible: (a) $g_k^{(z)}\left(P_\Omega^{(z)}, \tau\right) \approx g_{k-1}^{(y)}\left(P_\Omega^{(y)}, \tau\right)$, (b) $g_k^{(z)}\left(P_\Omega^{(z)}, \tau\right) > g_{k-1}^{(y)}\left(P_\Omega^{(y)}, \tau\right)$, and (c) $g_k^{(z)}\left(P_\Omega^{(z)}, \tau\right) < g_{k-1}^{(y)}\left(P_\Omega^{(y)}, \tau\right)$. In the first case, the polarization components of the XUV radiation are amplified with equal efficiency, which makes it possible to amplify elliptically or circularly polarized radiation while maintaining the degree of ellipticity. In case (b), $z$-polarized radiation is amplified more efficiently than $y$-polarized radiation. As a result, as the field propagates through the medium, the ellipticity of the field changes. If quasimonochromatic XUV radiation with low ellipticity and the major axis of the polarization ellipse parallel to the $y$ axis, which can be obtained via HHG process driven by an elliptically polarized laser field, enters the medium, then, as it is amplified in the medium, the ellipticity of the XUV radiation will increase, and at certain position inside the medium, its polarization will become close to circular. However, if the major axis of the polarization ellipse of the XUV radiation at the entrance to the medium coincides with the $z$ axis, its ellipticity will decrease during propagation in the medium, and the polarization of the output radiation will approach linear. In case (c), reasoning similar to case (b) holds up to 90° rotation of the polarization ellipse of the incident field. It should be noted that according to the solutions (14) and (16) derived above, both polarization components of the XUV radiation are amplified independently of each other, while the phase relations between them are preserved.

The half-sum and difference of gain factors for $z$- and $y$-polarized radiation, which are responsible for the total efficiency and anisotropy of harmonic amplification, versus the $k$ number are shown in Fig. 1 in [21]. As can be seen from that figure, the effective gain difference is basically an alternating function of $k$, but for -14 ≤ $k$ ≤ -6, the difference in gain is close to zero. Accordingly, the gain lines with *k* inside this region are most suitable for amplifying radiation of elliptical or circular polarization while maintaining its polarization state. Below we consider the case $k = -10$ for amplifying a single harmonic, and, similarly to [21], but in more details, $k = -13, -12, -11, -10$, and $-9$ for a set of 5 harmonics forming a sequence of short pulses of circular polarization. At the same time, for positive values of *k*, maximum gain anisotropy is achieved. Further, we will also consider the case $k = 10$, in which the gain for the *z*-polarized radiation is close to the maximum and the gain for the *y*-polarized radiation is almost zero. These conditions are optimal for enhancing the harmonic field ellipticity in the process of its amplification. In addition, similarly to [21], we will consider the case of amplification of a set of 5 high-order harmonics of elliptical polarization, which are in resonance with induced gain lines that correspond to $k = -3, -1, 1, 3$, and 5. For all the aforementioned gain lines, the *y*-polarized radiation is amplified more strongly than the *z*-polarized radiation, which allows one to consistently change the ellipticity of the set of high harmonics.

## IV. NUMERICAL RESULTS FOR THE AMPLIFICATION OF ELLIPTICALLY AND CIRCULARLY POLARIZED HIGH-ORDER HARMONICS

We will characterize the ellipticity of the XUV radiation by the quantity $\sigma$. In the case when the amplified radiation is elliptically polarized and the principal axes of the polarization ellipse coincide with the *y* and *z* axes, which will be considered throughout the paper, the following equality holds:

$$\frac{E_z^2}{\left|\tilde{E}_z\right|^2} + \frac{E_y^2}{\left|\tilde{E}_y\right|^2} = 1, \tag{18}$$

where $E_z$ and $E_y$ are the projections of the electric field vector of the XUV radiation on the *z* and *y* axes, respectively, and $\tilde{E}_z$ and $\tilde{E}_y$ are its slowly varying complex amplitudes. In this case, the radiation ellipticity is expressed as $\sigma = \sigma_c$, where the index "c" denotes "canonical" (in other words, intrinsic), and

$$\sigma_c = \frac{\left|\tilde{E}_z\right|^2}{\left|\tilde{E}_y\right|^2} \text{ if } \left|\tilde{E}_z\right|^2 \leq \left|\tilde{E}_y\right|^2, \text{ and } \sigma_c = \frac{\left|\tilde{E}_y\right|^2}{\left|\tilde{E}_z\right|^2} \text{ otherwise.} \tag{19}$$

In general case, when the principal axes of the polarization ellipse, $y_1$ and $z_1$, are rotated relative to the *y* and *z* axes by an arbitrary angle $\varphi$, see Fig. 4, the dependence of $\sigma$ on the values of $\tilde{E}_z$ and $\tilde{E}_y$ becomes more complex (see Appendix):

$$\sigma = \frac{(\sigma_c+1)/2 - \sqrt{[(\sigma_c+1)/2]^2 - \sigma_c \sin^2(\delta)}}{(\sigma_c+1)/2 + \sqrt{[(\sigma_c+1)/2]^2 - \sigma_c \sin^2(\delta)}}, \qquad (20)$$

where $\delta \equiv \arg(\tilde{E}_z) - \arg(\tilde{E}_y)$ is the phase difference between the slowly varying field amplitudes $\tilde{E}_z$ and $\tilde{E}_y$. Equation (20) reduces to (19) for $\delta = 0$ and $\delta = \pm n\pi/2$, where $n$ is an integer. We note that in the cases $\delta = 0$ or $\delta = \pi$, the radiation has linear polarization and $\sigma = 0$. In another limiting case, $\delta = \pm \pi/2$, when the condition $\tilde{E}_z^2 = \tilde{E}_y^2$ is satisfied, the radiation polarization is circular and $\sigma = 1$. In the general case, the ellipticity lies in the interval between zero and one, $0 \leq \sigma \leq 1$.

In what follows, we will consider the amplification of circularly and elliptically polarized XUV fields, both ($z$- and $y$-) polarization components of which are tuned to exact resonance with the corresponding gain lines. In this case, the quantities $g_k^{(z)}$ and $g_{k-1}^{(y)}$ are real, and if the principal axes of the polarization ellipse of the incident field coincide with the $y$ and $z$ axes, their orientation will not change during amplification. The ellipticity of the output radiation for this case is determined by a ratio of intensities of its polarization components and can be calculated using Eq. (19).

However, if the principal axes of the polarization ellipse of the incident field differ from the $y$ and $z$ axes, the general expression (20) will need to be used to calculate the ellipticity. Note that in the process of the harmonic amplification, it is possible to control not only their ellipticity, but also the angle of rotation of the principal axes of the polarization ellipse, $\varphi$ (see Fig. 5), which is expressed through the characteristics of the XUV radiation in the laboratory coordinate system as

$$\varphi = \frac{1}{2} \operatorname{arctg} \left( \frac{2|\tilde{E}_z||\tilde{E}_y|\cos(\delta)}{|\tilde{E}_z|^2 - |\tilde{E}_y|^2} \right). \qquad (21)$$

In particular, the control of the orientation of the principal axes of the polarization ellipse is possible if at least one of the polarization components of the field is detuned from the corresponding resonance (the condition (17) is not satisfied exactly). In this case, due to resonant dispersion, the corresponding gain will become complex, which will lead to a change in the phase difference between the $z$- and $y$- polarized components of the field as it propagates in the medium.

In the following, we consider an active medium with free-electron density $N_e = 5 \times 10^{19} \mathrm{cm}^{-3}$ and a density of neon-like $\mathrm{Ti}^{12+}$ ions $N_{ion} \approx 4.2 \times 10^{18}$ cm$^{-3}$. Among these ions approximately 1% are in the $|1\rangle$ state at $\tau=0$ (the initial population of the $|2\rangle-|4\rangle$ states is close to zero due to their rapid radiation depletion). In this case, the small signal gain (in intensity) for the resonant XUV radiation in the absence of modulation will be about $G_{total}^{(0)} = 70 \, \mathrm{cm}^{-1}$. The collision rate, evaluated from the experimentally measured spectral width of the gain line of an optically thin medium, $\Delta\lambda/\lambda = 1.5 \times 10^{-4}$, is

$1/\gamma_{Coll} = 213$ fs. The times of tunneling ionization from the |1⟩-|4⟩ states under the action of a modulating field with an intensity of $8.26\times10^{16}$ W/cm$^2$ are $1/w_{ion}^{(1)} \approx 124$ ps, $1/w_{ion}^{(2)} \approx 296$ ns, and $w_{ion}^{(3)} = w_{ion}^{(4)} \approx 45$ μs.

In contrast to the analytical approach, which assumed a step-like envelope (12) for the XUV radiation to be amplified, numerical calculations are performed for the case of a smooth envelope given by

$$\tilde{E}_z^{(inc)}(\tau) = E_0^{(z)}\left[\theta(\tau)-\theta(\tau-\tau_{zero})\right]\sin^2\left(\pi\tau/\tau_{zero}\right),$$
$$\tilde{E}_y^{(inc)}(\tau) = E_0^{(y)}\left[\theta(\tau)-\theta(\tau-\tau_{zero})\right]\sin^2\left(\pi\tau/\tau_{zero}\right),$$
(22)

where the parameter $\tau_{zero}$ determines the duration of the seed field envelope from zero to zero. In further calculations, we assume $\tau_{zero} = 750$ fs. In this case, the FWHM of intensity envelope is about 270 fs. These values of the parameters were chosen from the consideration that for effective amplification of the incident field, its duration should exceed the transient time for resonant polarization of the active medium, which in our case is about 200 fs. To amplify shorter pulses, it is necessary to shorten the polarization transient time by (a) increasing the collisional relaxation rate $\gamma_{Coll}$ (using a denser and/or hotter plasma [45, 46]), or (b) increasing the ionization rates $w_{ion}^{(i)}$, $i = $ 1,2,3,4 (using a more intense modulating field). In both cases, however, the broadening of the gain spectrum of the active medium will be accompanied by a decrease in its peak value. Another possibility for amplifying short pulses is to use high-intensity seed radiation, in the presence of which relaxation processes in the active medium can be neglected (see the section in [20] where the separation of a single attosecond pulse from a train is addressed). However, the consideration of this amplification regime is beyond the scope of this article. Further, we will consider the case of a seeding radiation with the intensity of polarization components in the range of $10^6$-$10^8$W/cm$^2$; in this case, for the considered lengths of the medium (up to 15 mm), the amplification of radiation occurs in a regime close to linear.

### A. Ellipticity-increasing amplification of a single harmonic

Here we consider the amplification of the 175th harmonic ($\lambda_{H175}$=22.29 nm) of the modulating field; this harmonic is resonant to the induced gain lines with $k$=10 (for $z$-polarization) and ($k$ - 1)=9 (for $y$-polarization). As already mentioned, in this case, the gain for the $z$-component of the XUV radiation is close to maximum, whereas the gain for the $y$-component is minimum, which makes it possible to effectively control the ellipticity of the radiation during its amplification.

Top panel of Fig. 5 shows the dependence of the amplified radiation ellipticity (19) on the local time $\tau$ and the spatial coordinate $x$ (length of the active medium). The intensities of the $y$- and $z$- components of the seed radiation are assumed to be $10^7$ W/cm$^2$ and $10^6$ W/cm$^2$, respectively. Accordingly, at $x = 0$ and $0 \leq \tau \leq 750$ fs, the radiation ellipticity is 0.1, and the y-polarization is dominant (at later times, the incident field is zero). As the radiation propagates in the medium, its ellipticity increases as a result of the predominant amplification of the z-polarized component, but this growth occurs nonuniformly in time, see Fig. 6, which shows the intensity time dependences (a), as well as the amplitude spectra (b) of the polarization components of the XUV field at different lengths of the medium. The amplification of the z-polarized field proceeds in a regime close to linear (without a significant change of the population difference in the medium) and is accompanied by a narrowing of its spectrum, see Fig. 6(b), as well as a lengthening of its envelope, see Fig. 6(a). Thus, for an arbitrary length of the medium, a time interval appears, during which the z-polarized radiation component becomes more intense than its y-polarized counterpart, so that there is a moment of time, at which the ellipticity of radiation approaches unity, see the top panel of Fig. 5 and the inset in Fig. 6(a), showing the cuts of Fig. 5 for a few fixed propagation distances. For small lengths, the z-polarized component dominates at the very end of the pulse (near $\tau = 750$ fs), where the y-polarized field tends to zero, whereas the z-component is nonzero due to amplification. As the length of the medium increases, the z-polarized radiation becomes comparable in amplitude with the y-polarized one at earlier moments of the local time. As a result, the area of maximum ellipticity shifts to the maximum of the field envelope. It should be noted that the maxima of the envelopes of the z- and y-polarized fields coincide only at $x = 0$, since in the process of amplification, the maximum of the envelope of the z-component shifts towards later moments of local time. The spatial dependence of the positions of the maxima of the z- and y- field components is shown on the top panel in Fig. 5 with green dashed and red dash-dotted curves. In addition, the orange solid curve in Fig. 5 shows the position of the maximum of the total intensity of the polarization components: in a thin medium, the maximum of the total intensity is achieved near the maximum of the y-polarized field envelope (z-polarized component is weaker), while at the maximum considered medium length, the maximum of the total intensity shifts to the maximum of the envelope of z-polarized field, which becomes dominant due to its strengthening discussed above. The bottom panel of Fig. 5 shows the spatial dependence of the field ellipticity at the time instant corresponding to the maximum of the total intensity of its polarization components. As can be seen, with an increase of the medium length, the field ellipticity increases due to an increase of the intensity of the z-polarized component. At $x \approx 5.7$ mm, the radiation at the maximum of the envelope of the total intensity of both polarization components acquire polarization approaching circular, $\sigma \simeq 1$, while for larger lengths of the medium, the output ellipticity decreases because of a further increase of the intensity of the z-polarized field. At the optimal length of the medium, $x = 5.7$ mm, the peak intensities of the z- and y-components of the field are approximately equal, but the energy contained in the z-component turns out to be much larger due to an increase in its duration during amplification. We also note that for the considered intensity of the incident field (~ $10^7$ W/cm$^2$), the energy of the amplified spontaneous emission from the active medium is insignificantly small as compared to the amplified HH signal. The only manifestation of the ASE is a change of ellipticity of the radiation at the very initial moments of time, $\tau \leq 50$ fs (before the HH signal becomes noticeable), see the top panel of Fig. 5 and the inset in Fig. 6(a). Figures 5 and 6 demonstrate that the use of a modulated active medium of a plasma-based X-ray laser makes it possible to increase the ellipticity of a single harmonic by an order of magnitude with a simultaneous increase in the radiation energy by nearly three times.

### B. Polarization-maintaining amplification of a circularly polarized single harmonic

We will now consider the amplification of the 135th harmonic ($\lambda_{H135}$=28.89 nm) of the modulating field; this harmonic is resonant to the induced gain lines with $k = -10$ (for $z$-polarization) and $(k - 1) = -11$ (for $y$-polarization). In this case, the gain factors for different polarization components are close to each other (see Fig. 4), which makes it possible to amplify the circularly or elliptically polarized radiation without a noticeable change in its polarization state. We will consider the case of a circularly polarized field ($\sigma$=1); at the entrance to the medium, the peak intensities of the $z$- and $y$-polarized components coincide and are equal to $10^7$ W/cm$^2$. The space-time dependence of the ellipticity of the amplified radiation is shown in Fig. 7 (top panel), where the green solid curve also marks the position of the maximum of the radiation intensity envelope (the maxima of the envelopes of different polarization components coincide on the scale of the figure). The bottom panel of Fig. 7 shows the field ellipticity at the maximum of its envelope versus the length of the medium. As follows from Fig. 7, the field ellipticity decreases monotonically with an increase of the medium length due to a slightly different amplification of its polarization components (the $z$-component is amplified somewhat more strongly). Nevertheless, even for the maximum considered length of the medium, the radiation retains its polarization close to circular: $\sigma \approx 0.89$ for $x$=10 mm. The time dependences of the intensities of the polarization components of the field and the corresponding spectra at different lengths of the medium are shown in Fig. 8, panels (a) and (b), respectively. As can be seen from Fig. 8(a), in the course of amplification, the field peak intensity increases approximately 27 times, while its energy increases approximately 62 times. The latter is associated with the lengthening of the field envelope. Accordingly, in the process of amplification, the emission spectrum narrows. Similarly to the case of ellipticity-increasing amplification, the energy of ASE is insignificant (however, ASE manifests itself in a distortion of the ellipticity space-time dependence at the front edge of amplified HH signal, $\tau \leq 50$ fs). Thus, the proposed approach makes it possible to boost the energy and peak intensity of circularly polarized resonant radiation by tens of times at the cost of the ellipticity decrease by approximately 10% (the ellipticity is less reduced at the leading edge of the XUV pulse, and decreases more at its trailing edge).

### C. Polarization-maintaining amplification of a set of high harmonics

Next, we will consider the amplification of a set of five elliptically or circularly polarized high-order harmonics forming a train of approximately 1 fs pulses. We will assume that at the entrance to the medium, each of the harmonics is characterized by an envelope of the form (22); while the initial phases, amplitudes, and polarization states of all harmonics are the same.

First, we will consider the amplification of a train of circularly polarized pulses ($\sigma=1$ at $x=0$) formed by a combination of 129th, 131st, 133rd, 135th, and 137th harmonics of the modulating field ($k = -13, -12, -11, -10$, and $-9$, respectively). The pulse duration at the entrance to the medium is 1.2 fs, the pulse repetition period is 6.5 fs, the central wavelength is 29.33 nm, and the peak intensity of each of the $y$- and $z$-polarization components of the harmonic field is $5\times10^7$ W/cm$^2$. It is worth noting that using shorter wavelength of the modulating field allows one to consider the amplification of proportionally shorter pulses. Thus if, instead of the considered 3.9 μm modulating field, one would use the modulating field with intensity $1.19\times10^{17}$ W/cm$^2$ and wavelength 2.7 μm, the duration and repetition period of the pulses will be 0.83 fs and 4.5 fs, respectively. The space-time dependence of the ellipticity of the amplified radiation, as well as the position of the maximum of the envelope of its intensity (the envelopes of the polarization components almost coincide on the scale of the figure), are shown on the top panel in Fig. 9. The bottom panel of Fig. 9 shows the spatial dependences of the ellipticity of each of the harmonics separately at the time instant corresponding to the maximum of the intensity envelope of the total harmonic field (which itself depends on the propagation distance, see the green curve on the top panel of Fig. 9). As seen from Fig. 9, the ellipticity of harmonics of different orders decreases at different rates, which are determined by the differences between the gain factors for the $y$- and $z$- components of the field of each harmonic, see Fig. 3 and also Fig. 1 in [21]. In general, Fig. 9 resembles Fig. 7, however, in contrast to the latter, in Fig. 9, a small-scale structure appears due to a change in the contribution of each of the harmonics to the resulting field on a time scale of a fraction of the IR field cycle. This structure in the function $\sigma(x,\tau)$ is shown in Fig. 10, which is plotted for a time interval equal to a half-cycle of the modulating field and located in the region of the maximum of the amplified radiation envelope at $x = 10$ mm. For clarity, Fig. 10 also shows the time dependence of the intensity of the seed radiation (the evolution of the pulse shape with increasing length of the medium is illustrated in Fig. 11(a)). As follows from Fig. 10, at the moments of time corresponding to the constructive interference of harmonics and the formation of a burst in the intensity of the resulting field, the ellipticity also reaches its maximum, which is $\sigma \approx 0.89$ at $x =10$ mm. The ellipticity decreases significantly only where harmonics of different orders are in antiphase with each other, and the resulting field is close to zero. The main features of the small-scale structure of the function σ($x,\tau$) shown in Fig. 10 are repeated on every half-cycle of the modulating field in the most energetic part of the pulse train.

In Fig. 11 we show the local time dependences of intensity (a) and ellipticity (b) of the XUV radiation within the same sub-IR-field-cycle time interval at a few propagation distances through the medium. As follows from Fig. 11(a), with increasing propagation distance the pulse duration increases from 1.2 fs to 1.6 fs because of nonuniform amplification of the harmonics of different orders (see Fig. 12(b)), while the peak ellipticity decreases from 1 to 0.89 because of a slightly stronger amplification of $z$-polarization component of the HH field as compared to the $y$-polarization, see Fig. 12(a). The appearance of the dips in the ellipticity time dependence in Fig. 11(b) is caused by the difference in the timing of the minima of the $z$- and $y$-polarization components of the field, which, in turn, originates from the difference in the spectral composition of the polarization components; see Fig. 12(b).

In Fig. 12 we plot the picosecond-scale local time dependences of the intensities (a) and the amplitude spectra (b) of the polarization components of the XUV radiation at different lengths of the medium. Fig. 12(a) generally resembles Fig. 8(a), which corresponds to the polarization-maintaining amplification of a single harmonic. In turn, the output radiation spectra in Fig. 12(b) generally follow the gain spectrum of the active medium shown in Fig. 3: the 133rd and 135th harmonics corresponding to $k=$ -11 and $k=$ -10 are amplified most efficiently, while the rest of the harmonics are amplified to a lesser degree, which results in reduction of the effective width of the harmonic emission spectrum with increasing propagation distance. Similar to Fig. 8(b), the emission spectrum of each individual harmonic narrows and the duration of its envelope increases, which leads to an increase in the duration of the resulting pulse train in Fig. 12(a). Furthermore, the spectral composition of the radiation varies with time: in the vicinity of the maximum of the intensity envelope, harmonics of different orders have comparable amplitudes, while in the tail of the envelope, the 133rd and 135th harmonics, which experienced the greatest amplification, dominate. Figure 13 illustrates the results of amplification of the HH field in 10 mm long active medium. Panel (a) shows the picosecond scale time dependencies of intensity and the ellipticity envelope (which is the ellipticity at the maxima of the individual pulses) of the total harmonic field, superimposed on each other, while in panel (b) we plot the time dependencies of intensity and ellipticity within a half-IR-field-cycle at the peak of the intensity envelope in (a). As follows from Fig. 13(a), at a medium length of 10 mm, the peak total intensity of the radiation increases 11.2 times (to a value of about $1.12 \times 10^9$ W/cm$^2$), and due to an increase in the duration of the polarization components of the field, as well as an increase in the duration of each of the pulses in the train, the radiation energy increases 35.5 times. At the maximum of the intensity envelope (at $x$ = 10 mm and $\tau$ = 540.2 fs), the ellipticity of the amplified radiation is 0.89. According to Fig. 13(b), the ellipticity does not change considerably within the pulse duration.

Thus, using a modulated active medium of a plasma-based X-ray laser, it is possible to increase the energy of XUV pulses of circularly polarized radiation by more than an order of magnitude without significantly changing their polarization state.

### D. Ellipticity-increasing amplification of a set of high harmonics

Finally, we will consider the possibility of increasing the ellipticity of the train of sub-femtosecond pulses in the process of their amplification using the example when such a train is formed by a combination of the 149th, 153rd, 157th, 161st, and 165th harmonics of the modulating field ($k= -3, -1, 1, 3$, and 5, see Fig. 3 and also Fig. 1 in [21]). In this case, the pulse duration at the entrance to the medium is 590 as, the pulse repetition period is 3.25 fs, and the central wavelength is 24.85 nm. The peak intensity of the $y$-component of the total harmonic field at $x=0$ is $5\times10^7$ W/cm$^2$. We take the peak intensity of the $z$-component of the field to be $1.67\times10^8$ W/cm$^2$, and the ellipticity of the field at the entrance to the medium is $\sigma(x=0,\tau)=0.3$. The change of ellipticity due to the predominant amplification of the $y$-component of the field as it propagates in the medium is shown in Figs. 14 and 15. Figure 14 is drawn for a time interval exceeding the duration of the envelope of the amplified radiation, while in Fig. 15, the dependence $\sigma(x,\tau)$ is shown for the half-cycle of the modulating field in the vicinity of the maximum of the radiation envelope at $x=11$ mm (which is the optimal propagation distance, see the discussion below). Top panel of Fig. 14 also depicts the spatial dependences of the time instants at which the maximum of the harmonic intensity envelope is reached for each of the polarization components of the XUV field and for their sum. The bottom panel of Fig. 14 shows the spatial dependences of the ellipticity of each harmonic separately at the maximum of the envelope of the total harmonic field (the sum over the polarizations). In general, Fig. 14 is similar to Fig. 5 demonstrating the ellipticity-increasing amplification of a single harmonic. With an increase of the length of the medium, the maximum of harmonic ellipticity shifts from $\tau\approx750$ fs to earlier moments of local time, while the maximum of the intensity envelope, on the contrary, shifts from $\tau = 375$ fs to later times. In the vicinity of $x=11$ mm, these maxima coincide at $\tau\approx452$ fs; in this case, at the maximum of the total intensity of the polarization components, the harmonic field acquires almost circular polarization, $\sigma=0.995$. At the same time, similar to the case of polarization-maintaining amplification of five harmonics shown in Fig 9, a small-scale structure appears in the space-time dependence of the harmonic ellipticity, due to (i) the difference in the ellipticity of the harmonics of different orders and (ii) a change in the conditions of their interference on a time scale of a fraction of the IR field cycle. This structure in the function $\sigma(x,\tau)$ is shown in Fig. 15. Similar to Fig. 10, the maximum ellipticity of the XUV radiation coin-cides with the maximum of its intensity, while where the intensity of the total radiation of harmonics is close to zero, a decrease of ellipticity is observed.

The time dependences of the intensities of the polarization components of the amplified radiation and the corresponding spectra at different lengths of the medium are shown in Fig. 16, panels (a) and (b), respectively. Just as in the case of ellipticity-increasing amplification of a single harmonic (see Fig. 6), the polarization component experiencing the greatest amplification in the medium (in this case, $y$-component) is stretched in time. At the same time, similar to the case of amplification of the set of five harmonics with approximate maintaining of their polarization (see Fig. 12), as a result of the predominant amplification of the 153rd, 157th, and 161st harmonics of the modulating field, the effective width of the radiation spectrum decreases during amplification. It leads to lengthening of the pulses from 590 as to 660 as in the vicinity of the maximum of the intensity envelope, see Fig. 17(a), where the evolution of the pulse shape is traced with an increase of the propagation distance. The corresponding evolution of the sub-IR-field-cycle time dependence of ellipticity is shown in Fig. 17(b), which represents the cuts of Fig. 15 for a few fixed propagation distances.

Figure 18 plots the superimposed time dependencies of the total intensity (summed over the polarization components) and ellipticity of the amplified XUV field at the optimal length of the medium, $x$=11 mm, on the picosecond (a) and a sub-IR-field-cycle (b) time scales. In (a) and (b) we plot the ellipticity envelope (the ellipticity at the maxima of individual sub-femtosecond pulses) and the instantaneous ellipticity of the total harmonic field, respectively. As can be seen from Fig. 18(a), the increase in ellipticity is accompanied by 5.1 times increasing energy of the XUV field. In turn, as follows from Fig. 18(b), the ellipticity of the HH field shows just a moderate variation within the pulse duration. Thus, the proposed method makes it possible to change the polarization of the train of sub-femtosecond XUV pulses from elliptical to nearly circular with a simultaneous increase in their energy by several times.

## V. CONCLUSION

In the accompanying Letter [21], we show for the first time the possibility of amplifying a train of sub-femtosecond pulses of XUV radiation with elliptical or circular polarization, constituted by high-order harmonics of the IR laser field, with maintaining or increasing the ellipticity of the amplified XUV radiation. For these purposes, it is proposed to use the active medium of neon-like $Ti^{12+}$ plasma-based X-ray laser additionally irradiated by a linearly polarized IR laser field of fundamental frequency (a replica of the field used to generate harmonics).

In the present paper, we provide more detailed analysis of this problem. To this end, we derive an analytical theory and present the results of numerical studies of the amplification of both a single harmonic detuned from resonance by an even multiple of the modulating field frequency and a set of such harmonics, constituting a train of few femtosecond or sub-femtosecond pulses. In each case, we study both the possibilities for the amplification with approximate polarization maintenance and the amplification with an ellipticity enhancement. In the case of single harmonic amplification, we show that the polarization component of the field which experiences the stronger amplification, is spectrally narrowed and lengthened in time, which results in time variation of ellipticity of the amplified radiation. This time variation is rather moderate in the case of the polarization maintaining amplification and much more substantial for the amplification with an ellipticity enhancement (however, in the latter case, at the optimal length of the medium, the maxima of radiation intensity and ellipticity coincide). In the case of amplification of a set of harmonics, we also show that because of the difference in the gain coefficients for the harmonics of different orders, at the output from an optically deep active medium some of the harmonics dominate over the others due to stronger amplification, which results in narrowing of the total harmonic spectrum and an increase in the duration of each femto- or sub-femtosecond pulse. In addition, the difference in both the amplitudes and the polarization states of harmonics of different orders, caused by slightly different conditions of their amplification, results in the variation of ellipticity of the XUV radiation within a half-cycle of the fundamental frequency IR field. However, in the most energetic part of the pulse train, the ellipticity reaches the maximum value at the moments of time corresponding to the formation of a burst in the resulting field of harmonics, and decreases significantly only where the harmonics of different orders are in antiphase with each other, and the resulting field is close to zero. Accordingly, only a moderate variation of ellipticity occurs within the duration of each of the amplified pulses.

Similarly to [21], we analyze the possibilities for the amplification of a train of circularly polarized pulses with a central wavelength of 29.33 nm and individual pulse duration of 1.2 fs, produced by the HHs of an IR field with 3.9 μm wavelength, in neon-like $Ti^{12+}$ active medium of a plasma-based X-ray laser, dressed by the fundamental frequency IR field with intensity $8.26 \times 10^{16}$ W/cm$^2$, with an approximate maintenance of the polarization state of the harmonic pulse during its amplification. Thus, an increase in the energy of the pulse train by 35 times is achieved at the cost of an ellipticity decrease by 11%. We also consider the amplification of a train of pulses with a central wavelength of 24.85 nm and a duration of 590 as, formed by the harmonics of the same modulating IR field, with an increase in the ellipticity by more than 3 times at the peak of the intensity envelope of the HH radiation (which corresponds to the transformation of elliptically polarized field into the circularly polarized one), accompanied by an increase of the radiation energy by more than 5 times.

The proposed method can be extended to other neon-like ions. In addition, due to the similarity of the energy structure of nickel-like and neon-like ions, this method can also be generalized to the case of nickel-like active media (in particular, those based on $Mo^{14+}$ or $Ag^{19+}$ ions). In prospect, this opens up the possibility of amplifying the emission of harmonics of elliptical and circular polarization in shorter-wavelength spectral ranges. All in all, this approach might allow for a considerable increase in both the energy and ellipticity of few femtosecond and sub-femtosecond XUV radiation pulses for probing chiral and magnetic media.

**ACKNOWLEDGMENTS**

The numerical calculations presented in this article were carried out with the support from the Russian Science Foundation (RSF, Grant No. 19-72-00140). The analytical studies were supported by the Russian Foundation for Basic Research (RFBR, Grant No. 18-02-00924). V.A.A. acknowledges personal support from the Foundation for the Advancement of Theoretical Physics and Mathematics BASIS. O.K. appreciates the support by the National Science Foundation (Grant No. PHY-2012194).

**APPENDIX**

For an arbitrary phase difference between slowly varying amplitudes of the $z$- and $y$-components of the XUV radiation, $\tilde{E}_z$ and $\tilde{E}_y$, in a plane polarized monochromatic wave traveling along the $x$ axis, the end of the electric field vector draws, in general, an ellipse given by

$$\frac{E_z^2}{|\tilde{E}_z|^2} + \frac{E_y^2}{|\tilde{E}_y|^2} - 2\frac{E_z}{|\tilde{E}_z|}\frac{E_y}{|\tilde{E}_y|}\cos(\delta) = \sin^2(\delta). \tag{A1}$$

At the same time, in the own coordinate system ($x_1,y_1,z_1$), the polarization ellipse of the XUV radiation is written in a canonical form:

$$\frac{E_{z_1}^2}{|\tilde{E}_{z_1}|^2} + \frac{E_{y_1}^2}{|\tilde{E}_{y_1}|^2} = 1, \tag{A2}$$

where $E_{z_1}$ and $E_{y_1}$ are the polarization components of the high-frequency field, whereas $\tilde{E}_{z_1}$ and $\tilde{E}_{y_1}$ are its complex amplitudes in the rotated coordinate system ($x_1,y_1,z_1$), see Fig. 4. In the coordinate system ($x_1,y_1,z_1$), the ellipticity of the XUV radiation is determined by an equation similar to (18) and is equal to the ratio of the minor and major semi-axes of the polarization ellipse,

$$\sigma_c = \frac{\left|\tilde{E}_{z_1}\right|^2}{\left|\tilde{E}_{y_1}\right|^2} \text{ if } \left|\tilde{E}_{z_1}\right|^2 \leq \left|\tilde{E}_{y_1}\right|^2 \text{ and } \sigma_c = \frac{\left|\tilde{E}_{y_1}\right|^2}{\left|\tilde{E}_{z_1}\right|^2} \text{ in the opposite case.} \quad (A3)$$

To obtain an expression for the ellipticity of radiation in the coordinate system ($x_1, y_1, z_1$), we express the quantities $\tilde{E}_{z_1}$ and $\tilde{E}_{y_1}$ through $\tilde{E}_z$ and $\tilde{E}_y$. Projecting the basis vectors from one set to another, we obtain

$$\vec{E} = E_z \vec{z}_0 + E_y \vec{y}_0 = E_z \left(\cos(\varphi)\vec{z}_{01} - \sin(\varphi)\vec{y}_{01}\right) + E_y \left(\sin(\varphi)\vec{z}_{01} + \cos(\varphi)\vec{y}_{01}\right) =$$
$$= \vec{z}_{01}\left\{E_z \cos(\varphi) + E_y \sin(\varphi)\right\} + \vec{y}_{01}\left\{-E_z \sin(\varphi) + E_y \cos(\varphi)\right\}$$

and

$$\begin{cases} E_{z_1} = E_z \cos(\varphi) + E_y \sin(\varphi), \\ E_{y_1} = -E_z \sin(\varphi) + E_y \cos(\varphi). \end{cases} \quad (A4)$$

Substituting (A4) into (A3), after some transformations, we arrive at

$$\left\{\left|\tilde{E}_{z_1}\right|^2 \sin^2(\varphi) + \left|\tilde{E}_{y_1}\right|^2 \cos^2(\varphi)\right\} E_z^2 + \left\{\left|\tilde{E}_{z_1}\right|^2 \cos^2(\varphi) + \left|\tilde{E}_{y_1}\right|^2 \sin^2(\varphi)\right\} E_y^2 -$$
$$- 2\cos(\varphi)\sin(\varphi)\left\{\left|\tilde{E}_{z_1}\right|^2 - \left|\tilde{E}_{y_1}\right|^2\right\} E_z E_y = \left|\tilde{E}_{z_1}\right|^2 \left|\tilde{E}_{y_1}\right|^2. \quad (A5)$$

Comparing (A2) and (A5), we obtain the following system of equations:

$$\left|\tilde{E}_y\right|^2 = \left|\tilde{E}_{z_1}\right|^2 \sin^2(\varphi) + \left|\tilde{E}_{y_1}\right|^2 \cos^2(\varphi), \quad (A6a)$$

$$\left|\tilde{E}_z\right|^2 = \left|\tilde{E}_{z_1}\right|^2 \cos^2(\varphi) + \left|\tilde{E}_{y_1}\right|^2 \sin^2(\varphi), \quad (A6b)$$

$$\left|\tilde{E}_z\right|\left|\tilde{E}_y\right|\cos(\delta) = \cos(\varphi)\sin(\varphi)\left\{\left|\tilde{E}_{z_1}\right|^2 - \left|\tilde{E}_{y_1}\right|^2\right\}, \quad (A6c)$$

$$\left|\tilde{E}_z\right|^2\left|\tilde{E}_y\right|^2 \sin^2(\delta) = \left|\tilde{E}_{z_1}\right|^2 \left|\tilde{E}_{y_1}\right|^2. \quad (A6d)$$

Next, we combine the first two equations as follows: (A6a) + (A6b) and (A6a) - (A6b). As a result, the system of equations (A6) takes the form

$$\left|\tilde{E}_z\right|^2 + \left|\tilde{E}_y\right|^2 = \left|\tilde{E}_{z_1}\right|^2 + \left|\tilde{E}_{y_1}\right|^2, \quad (A7a)$$

$$\left|\tilde{E}_z\right|^2 - \left|\tilde{E}_y\right|^2 = \left\{\left|\tilde{E}_{z_1}\right|^2 - \left|\tilde{E}_{y_1}\right|^2\right\}\cos(2\varphi), \quad (A7b)$$

$$2\left|\tilde{E}_z\right|\left|\tilde{E}_y\right|\cos(\delta) = \left\{\left|\tilde{E}_{z_1}\right|^2 - \left|\tilde{E}_{y_1}\right|^2\right\}\sin(2\varphi), \quad (A7c)$$

$$\left|\tilde{E}_z\right|^2 \left|\tilde{E}_y\right|^2 \sin^2(\delta) = \left|\tilde{E}_{z_1}\right|^2 \left|\tilde{E}_{y_1}\right|^2. \quad (A7d)$$

It is easy to see that Eq. (A7d) follows from Eqs. (A7a) - (A7c). Next, we square (A7b) and (A7c), and then combine the resulting equations. As a result, we obtain a system of equations for unknown quantities $\left|\tilde{E}_{z_1}\right|^2$ and $\left|\tilde{E}_{y_1}\right|^2$:

$$\left|\tilde{E}_z\right|^2 + \left|\tilde{E}_y\right|^2 = \left|\tilde{E}_{z_1}\right|^2 + \left|\tilde{E}_{y_1}\right|^2, \tag{A8a}$$

$$\left(\left|\tilde{E}_z\right|^2 - \left|\tilde{E}_y\right|^2\right)^2 + 4\left|\tilde{E}_z\right|^2\left|\tilde{E}_y\right|^2 \cos^2(\delta) = \left(\left|\tilde{E}_{z_1}\right|^2 - \left|\tilde{E}_{y_1}\right|^2\right)^2. \tag{A8b}$$

Further, expressing one unknown quantity through another from (A8a) and substituting this into (A8b), we obtain the following pair of solutions:

$$\left(\tilde{E}_{z_1}^2\right)_{1,2} = \frac{\left|\tilde{E}_z\right|^2 + \left|\tilde{E}_y\right|^2}{2} \pm \sqrt{\left(\frac{\left|\tilde{E}_z\right|^2 + \left|\tilde{E}_y\right|^2}{2}\right)^2 - \left|\tilde{E}_z\right|^2\left|\tilde{E}_y\right|^2 \sin^2(\delta)}, \tag{A9a}$$

$$\left(\tilde{E}_{y_1}^2\right)_{1,2} = \frac{\left|\tilde{E}_z\right|^2 + \left|\tilde{E}_y\right|^2}{2} \mp \sqrt{\left(\frac{\left|\tilde{E}_z\right|^2 + \left|\tilde{E}_y\right|^2}{2}\right)^2 - \left|\tilde{E}_z\right|^2\left|\tilde{E}_y\right|^2 \sin^2(\delta)}, \tag{A9b}$$

where the indices 1 and 2 denote the solutions corresponding to the upper and lower sign, respectively. This pair of solutions corresponds to two ellipses with the same major and minor axes, rotated relative to each other. Thus, for both solutions, the ellipticity is the same and equal to

$$\sigma = \frac{(\sigma_c + 1)/2 - \sqrt{\left[(\sigma_c + 1)/2\right]^2 - \sigma_c \sin^2(\delta)}}{(\sigma_c + 1)/2 + \sqrt{\left[(\sigma_c + 1)/2\right]^2 - \sigma_c \sin^2(\delta)}}, \tag{A10}$$

where $\sigma_c$ is determined by Eq. (19).

We also note that, in accordance with Eqs. (A7b) and (A7c), the rotation angle of the main axes of the polarization ellipse $y_1, z_1$ with respect to the $y$ and $z$ axes is expressed as

$$\varphi = \frac{1}{2}\operatorname{arctg}\left(\frac{2\left|\tilde{E}_z\right|\left|\tilde{E}_y\right|\cos(\delta)}{\left|\tilde{E}_z\right|^2 - \left|\tilde{E}_y\right|^2}\right). \tag{A11}$$

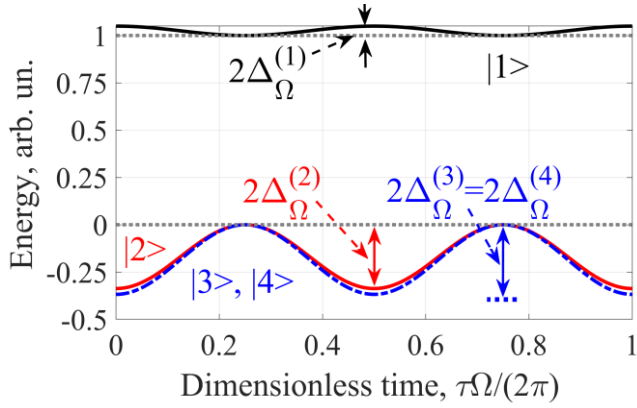

FIG. 1. Stark shift of the relevant energy levels of $Ti^{12+}$ ions calculated according to Eq. (4) under the action of the IR field with intensity $8.26\times10^{16}$W/cm$^2$, which is assumed in Figs. 3, 5-18. Black and red solid curves correspond to the states $|1\rangle=|3p\,^1S_0, J=0, M=0\rangle$ and $|2\rangle=|3s\,^1P_1, J=1, M=0\rangle$, respectively. Blue dash-dotted curve shows the degenerate energy level, which corresponds to the states $|3\rangle=|3s\,^1P_1, J=1, M=1\rangle$ and $|4\rangle=|3s\,^1P_1, J=1, M=-1\rangle$. Grey dotted horizontal lines are the position of the upper and lower lasing energy levels in the absence of the IR field.

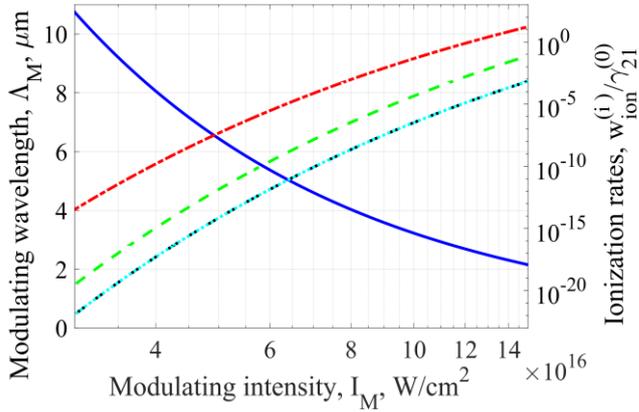

FIG. 2. Dependences of (i) the wavelength of the modulating field for which the condition (17) is satisfied (blue solid curve, left vertical axis) and (ii) ionization rates from the resonant states of $Ti^{12+}$ ions normalized to the half-width of the transition line in the absence of a field (other curves, right vertical axis) on the intensity of the modulating field. Red dash-dotted curve corresponds to the ionization rate from the $|1\rangle$ state, green dashed curve – from the $|2\rangle$ state, and the identical black and turquoise dashed curves – to the ionization rates from the $|3\rangle$ and $|4\rangle$ states.

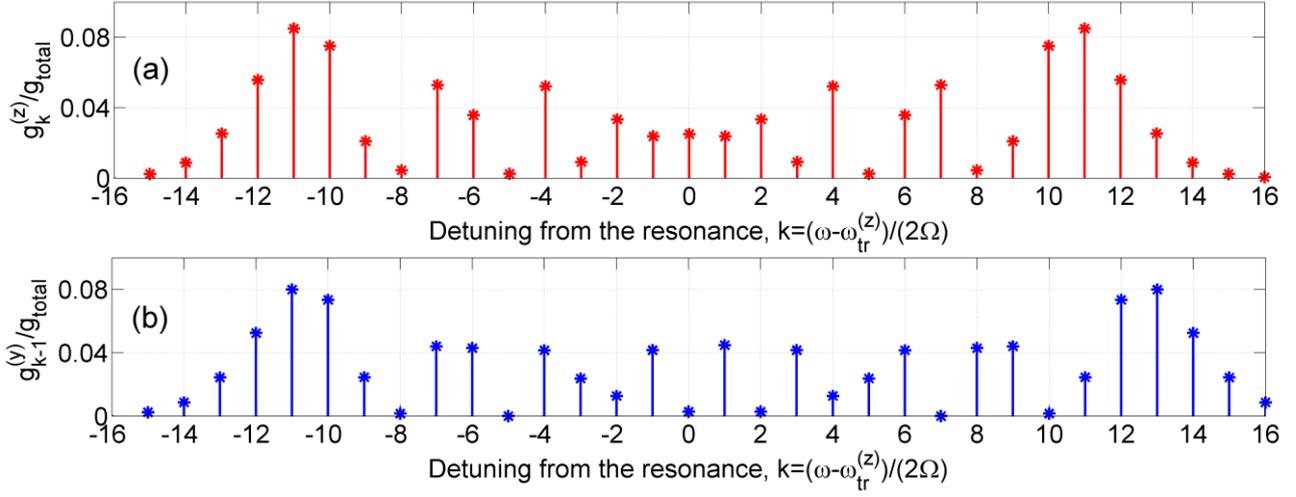

FIG. 3. Dependences of the effective gain coefficients (14b) and (16b) for the XUV radiation of $z$-polarization (a) and $y$-polarization (b), on the $k$ number, which corresponds to the detuning of the radiation frequency from resonance with a time-average $|1\rangle \leftrightarrow |2\rangle$ transition frequency. Effective gain factors are normalized to the gain in the absence of modulation, $g_{total}$. Here $P_\Omega^{(z)} \approx 12.57$, $P_\Omega^{(y)} \approx 13.57$, and $\tau \gg \gamma_z^{-1}, \gamma_y^{-1}$.

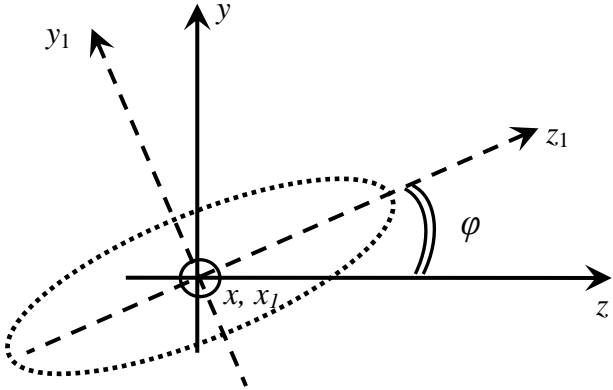

FIG. 4. Laboratory coordinate system $(x,y,z)$ and coordinate system $(x_1,y_1,z_1)$ in which the polarization ellipse of the XUV radiation takes the canonical form. The axis $x_1=x$ is directed into a page, $z_1 = z\cos(\varphi) + y\sin(\varphi)$, $y_1 = -z\sin(\varphi) + y\cos(\varphi)$.

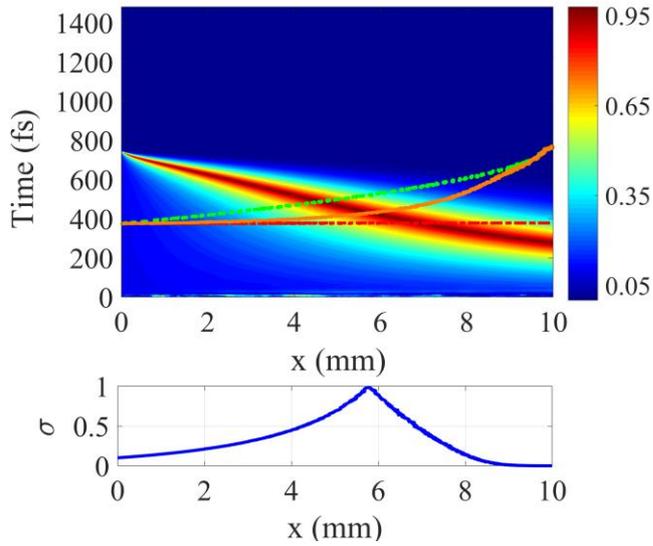

FIG. 5. Top panel: Space-time dependence of ellipticity, σ, of 175th harmonic of the modulating field ($k = 10$). At the entrance to the medium, $\sigma(x=0)=0.1$ (y-polarization dominates). Green dashed and red dash-dotted curves show the positions of the envelope maxima of the z- and y-polarization components of the amplified radiation, respectively. Orange solid curve shows the maximum of the total intensity of the polarization components. Some distortions in the very initial times, $\tau \leq 50$ fs, are caused by the ASE of the active medium. Bottom panel: Ellipticity of the XUV radiation at the maximum of the total intensity of its polarization components as a function of the length of the medium, $x$.

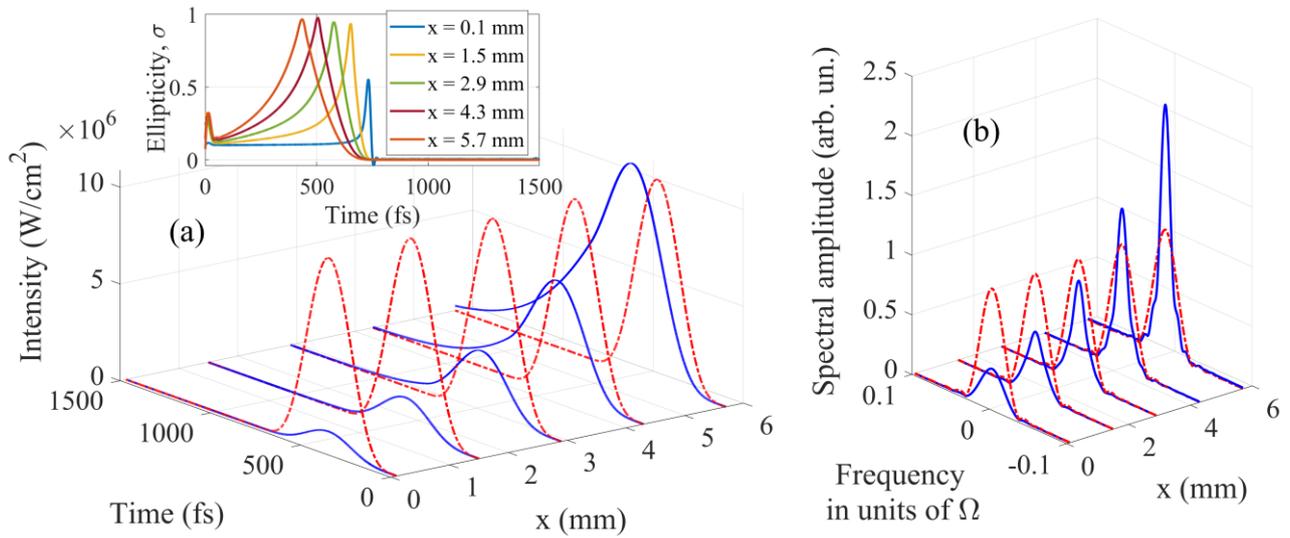

FIG. 6. (a) Dependences of the intensities of the polarization components of the XUV field on the local time, $\tau$, corresponding to Fig. 5, at different lengths of the medium: $x=0.1$ mm, 1.5 mm, 2.9 mm, 4.3 mm, and 5.7 mm. Red dash-dotted curve corresponds to the y-polarization component of the field, while blue solid curve shows the z-component. The inset shows the time dependences of ellipticity of the field at the same lengths of the medium. For a larger length, the peak of ellipticity shifts to a smaller time. The ellipticity increase at the very initial times is caused by the ASE. (b) Spectral amplitudes of the polarization components of the XUV field at the same propagation distances through the medium. Red dash-dotted curve and lower gain correspond to the y-component

of the field; blue solid curve and higher gain correspond to the *z*-component. The curves are normalized to the peak amplitude of the Fourier transform of the *y*-component of the seeding radiation.

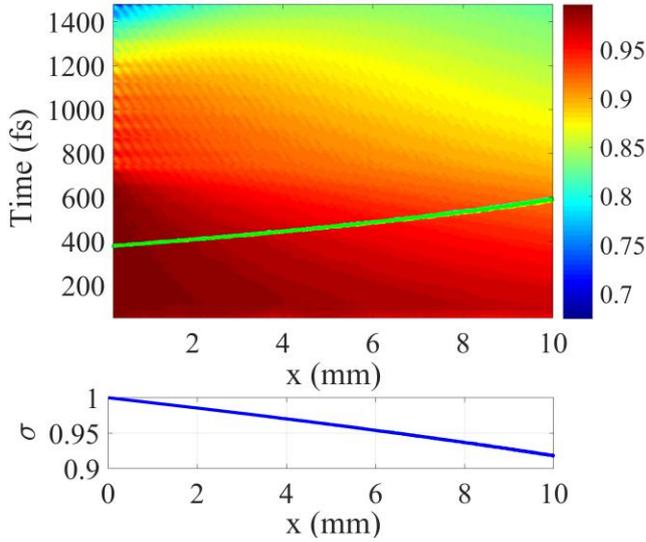

FIG. 7. Top panel: Space-time dependence of ellipticity, σ, of 135th harmonic of the modulating field (k = -10). The incident radiation is circularly polarized, σ(*x*=0) = 1. Green solid curve shows the positions of the maxima of the envelopes of the *z*- and *y*-polarized components of the field (they coincide on the scale of the figure). For the visibility of the figure, we don't show the interval $0 \leq \tau \leq 50$ fs, where the ASE dominates over the amplified signal. Bottom panel: Ellipticity of XUV radiation at the maximum of the total intensity of its polarization components as a function of the length of the medium, *x*.

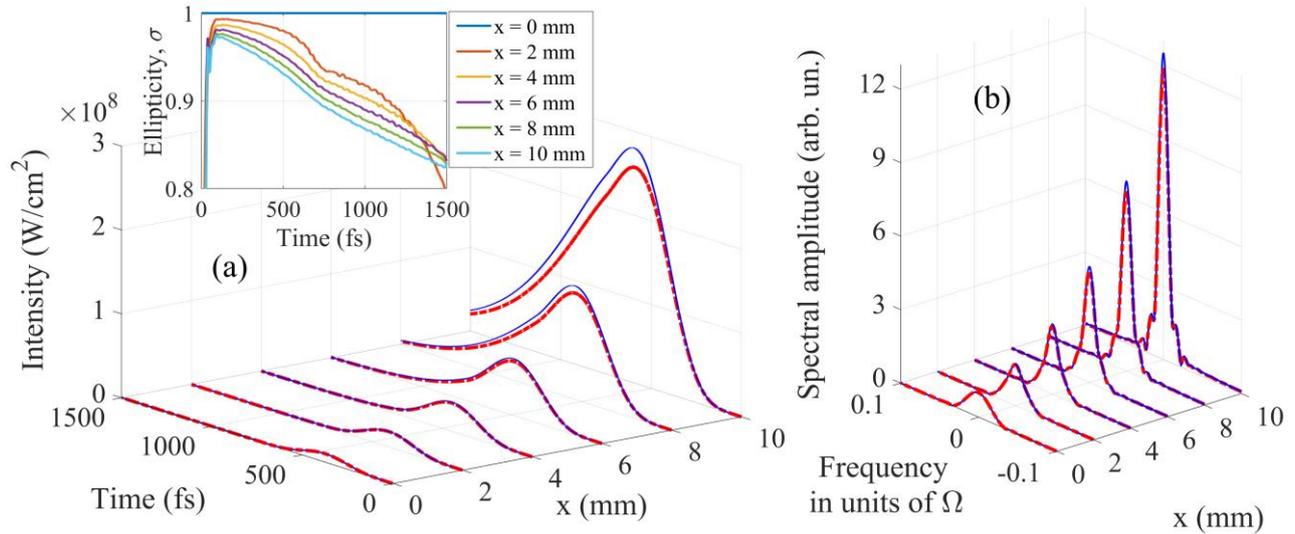

FIG. 8. (a) Dependences of the intensities of the polarization components of the XUV field on the local time, τ, corresponding to Fig. 7, at different lengths of the medium: *x*=0 mm (incident field), 2 mm, 4 mm, 6 mm, 8 mm, and 10 mm. Red dash-dotted curve corresponds to the *y*-polarization component of the field, while blue solid curve shows the *z*-component. The inset shows the time dependences of ellipticity of the field at the same lengths of the medium. With increasing medium length, the ellipticity decreases. The drop of ellipticity at $\tau \leq 50$ fs is caused by the ASE of the me-

dium. (b) Spectral amplitudes of the polarization components of the XUV field at the same propagation distances through the medium. Red dash-dotted curve and lower gain correspond to the *y*-component of the field; blue solid curve and higher gain correspond to the *z*-component. The curves are normalized to the peak amplitude of the Fourier transform of the *y*-component of the seeding radiation.

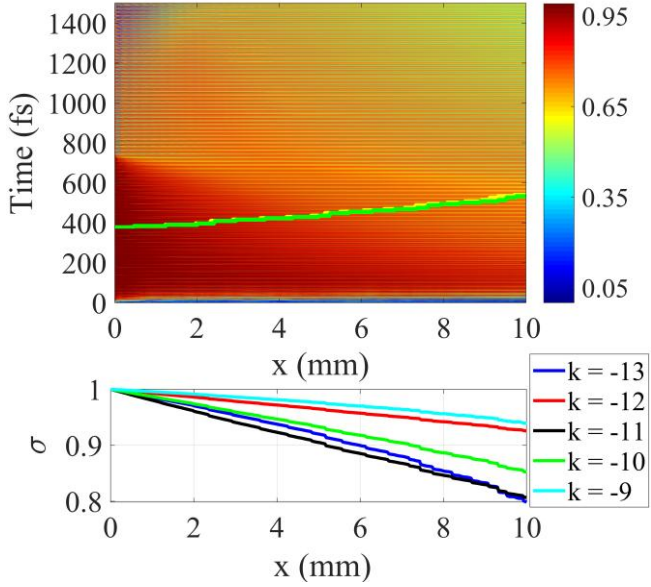

FIG. 9. Top panel: Space-time dependence of the ellipticity, $\sigma$, of the set of the 129th, 131st, 133rd, 135th, and 137th harmonics of the modulating field ($k = -13, -12, -11, -10$ and $-9$, respectively). The incident radiation is circularly polarized, $\sigma(x=0)= 1$. Largely overlapping green and yellow curves show the positions of the maxima of the envelopes of the *z*- and *y*-polarization components of the field, respectively. Bottom panel: Ellipticity of each individual harmonic (harmonics are numbered by the index *k*) at the maximum of the total intensity of all harmonics as a function of the length of the medium, *x*.

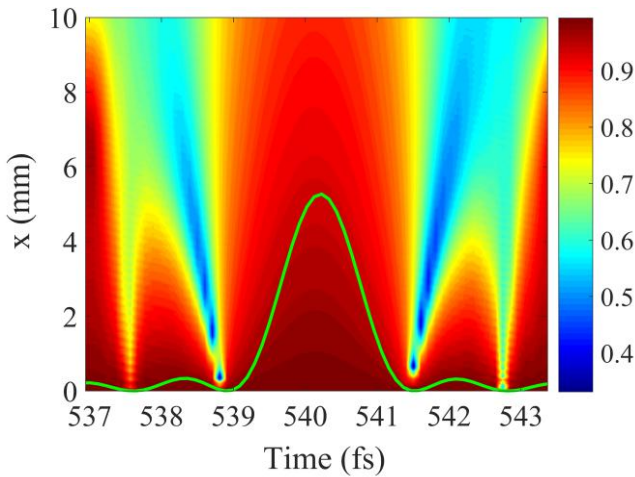

FIG. 10. Enlarged view of the fragment of Fig. 9 corresponding to the time interval 537 fs $\leq \tau \leq$ 543.4 fs equal in duration to the half-cycle of the modulating field. The vertical and horizontal axes are reversed. Green curve shows the shape of the pulses of the amplified radiation (time dependence of the intensity) at the entrance to the medium, *x*=0.

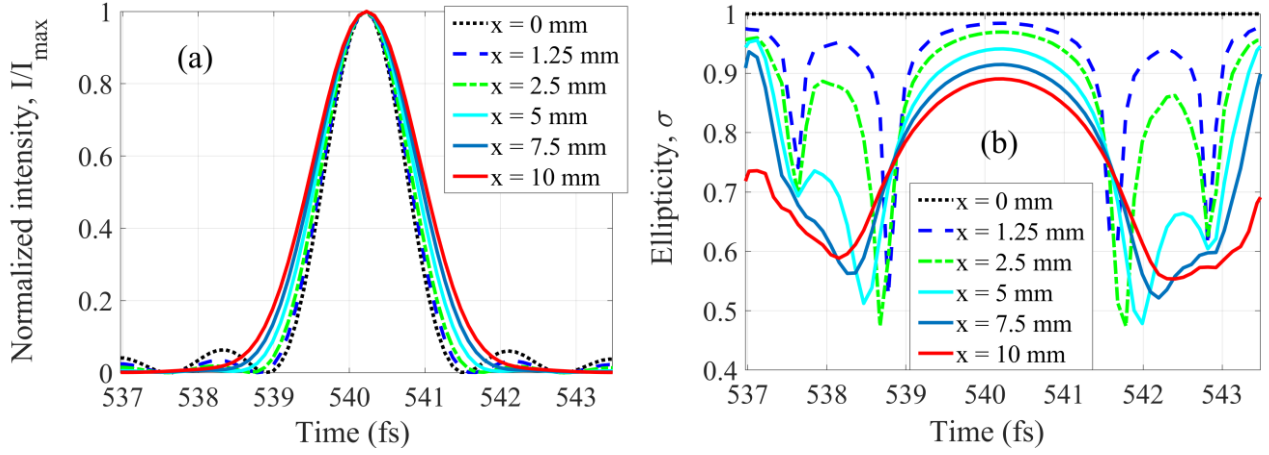

FIG. 11. (a) Local time dependence of the total intensity of the amplified XUV radiation corresponding to Fig. 10 within a half-cycle of the IR field at different lengths of the medium: $x=0$ mm (incident field), 1.25 mm, 2.5 mm, 5 mm, 7.5 mm, and 10 mm. The time interval is chosen at the peak of the envelope of the total intensity at $x=10$ mm, see Fig. 13. With increasing length of the medium, the pulse duration grows from $\tau_{pulse}=1.2$ fs at $x=0$ mm to $\tau_{pulse}=1.6$ fs at $x=10$ mm. (b) Corresponding variation of ellipticity of the XUV field within a half IR field cycle. With increasing medium length, the peak ellipticity decreases from $\sigma= 1$ at $x=0$ mm to $\sigma= 0.89$ at $x=10$ mm.

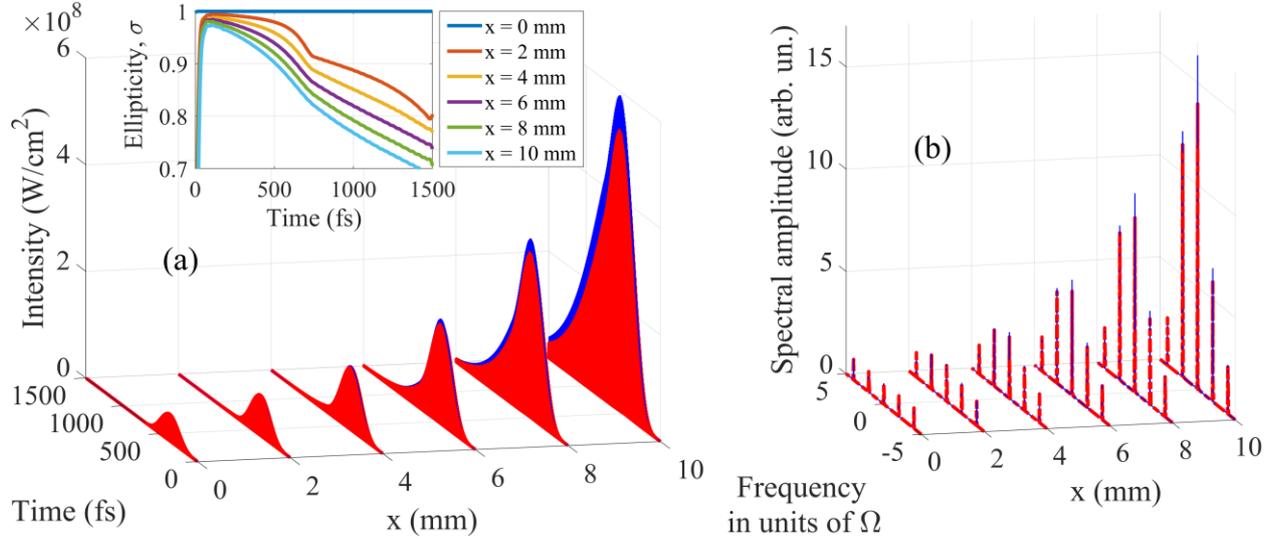

FIG. 12. (a) Dependences of the intensities of the polarization components of the XUV field on the local time, $\tau$, corresponding to Fig.9, at different lengths of the medium: $x=0$ mm (incident field), 2 mm, 4 mm, 6 mm, 8 mm, and 10 mm. Red curve corresponds to the $y$-polarization component of the field, while blue curve shows the $z$-component. The inset shows the time dependences of the ellipticity envelope of the field at the same lengths of the medium. With increasing length of the medium, the ellipticity decreases. The drop of ellipticity at $\tau \leq 50$ fs is caused by the ASE. (b) Spectral amplitudes of the polarization components of the XUV field at the same propagation distances in the medium. Red dash-dotted curve and lower gain correspond to the $y$-component of the field; blue thin solid curve and higher gain correspond to the $z$-component. The curves are normalized to the peak amplitude of the Fourier transform of the $y$-component of the seeding radiation.

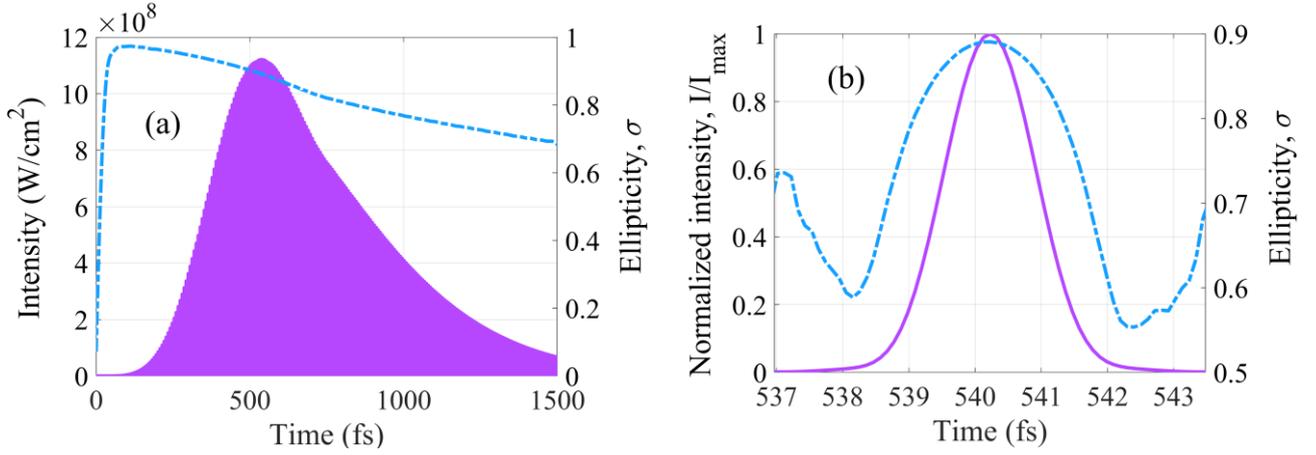

FIG. 13. (a) Local time dependences of the total intensity (solid lavender curve, left vertical axis) and the ellipticity envelope, see the text, (dash-dotted blue curve, right vertical axis) of the XUV pulse train, corresponding to Figs. 9-12, amplified in 10 mm thick medium. (b) Same as in (a), but within a single half-cycle of the IR field at the maximum of the intensity envelope (in this case, the dash-dotted blue curve shows the instantaneous value of the ellipticity). The XUV intensity in (b) is normalized to its peak value, $I_{max}=1.1\times10^9$ W/cm$^2$.

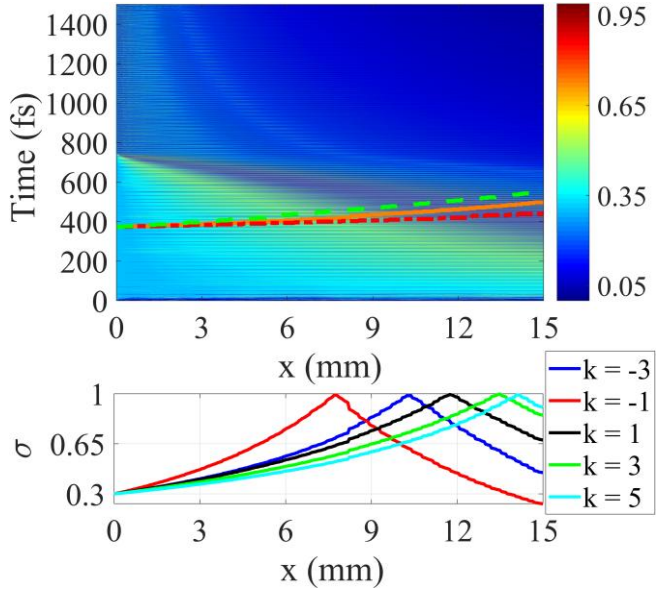

FIG. 14. Top panel: Space-time dependence of ellipticity, $\sigma$, of the set of the 149th, 153rd, 157th, 161st, and 165th harmonics of the modulating field ($k = -3, -1, 1, 3,$ and 5, respectively). At the entrance to the medium, $\sigma(x=0) = 0.1$ ($z$-polarization dominates). Red dash-dotted and green dashed curves show the positions of the maxima of the envelopes of the $z$- and $y$-polarized components of the total harmonic field. Orange solid curve shows the maximum of the total intensity of both polarization components. Bottom panel: Ellipticity of each individual harmonic (harmonics are numbered by the index $k$) at the maximum of the total intensity (summed over all harmonics and both polarization components) as a function of the medium length, $x$.

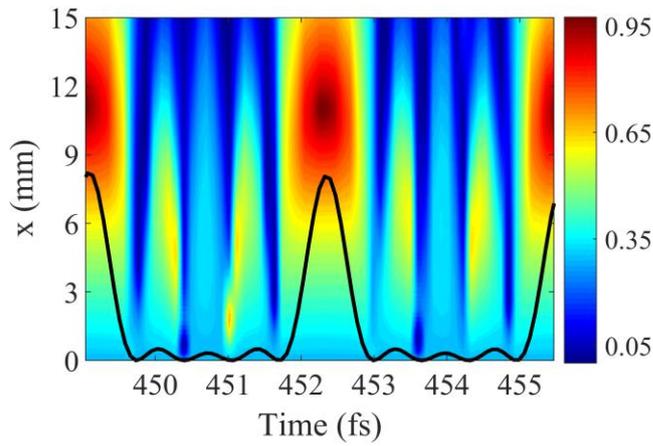

FIG. 15. Enlarged view of the fragment of Fig. 14 corresponding to the time interval 449 fs $\leq \tau \leq$ 455.5 fs. The vertical and horizontal axes are reversed. Black curve shows the shape of the pulses of the amplified radiation (time dependence of the intensity) at the entrance to the medium, $x=0$.

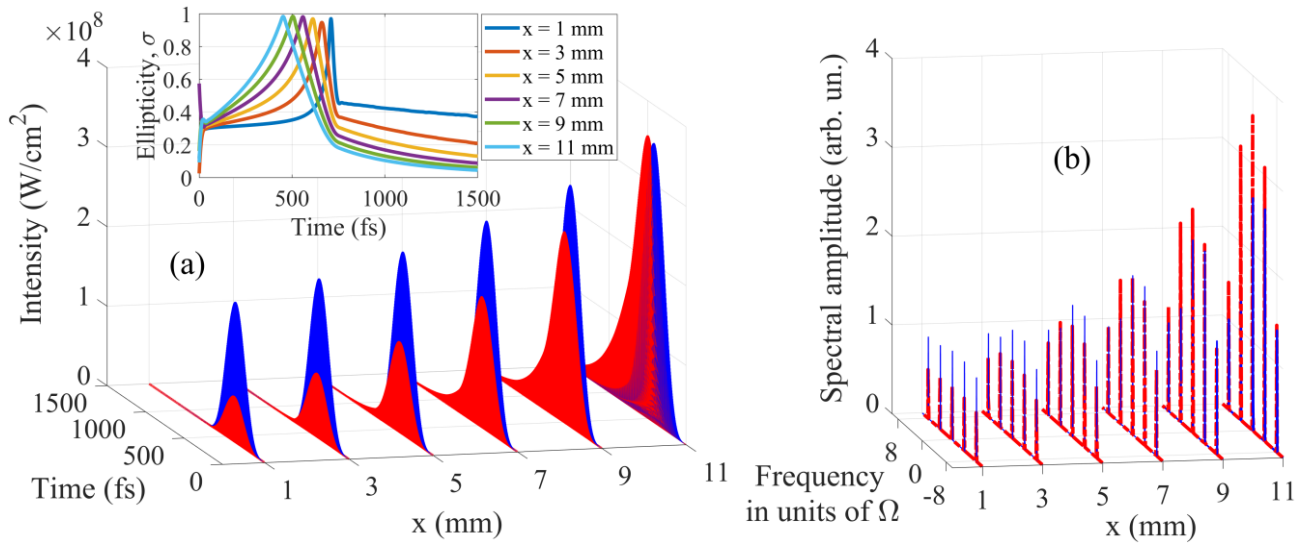

FIG. 16. (a) Dependences of the intensities of the polarization components of the XUV field on the local time, $\tau$, corresponding to Fig. 14, at different lengths of the medium: $x=1$ mm, 3 mm, 5 mm, 7 mm, 9 mm, and 11 mm. Red curve corresponds to the $y$-polarization component of the field, while blue curve shows the $z$-component. The inset shows the time dependences of the ellipticity envelope of the field at the same lengths of the medium. For a larger length, the peak of ellipticity shifts to a smaller time. The ellipticity increase at the very initial times is caused by the ASE. (b) Spectral amplitudes of the polarization components of the XUV field at the same propagation distances in the medium. Red dash-dotted curve and lower gain correspond to the $y$-component of the field; blue thin solid curve and higher gain correspond to the $z$-component. The curves are normalized to the peak amplitude of the Fourier transform of the $y$-component of the seeding radiation.

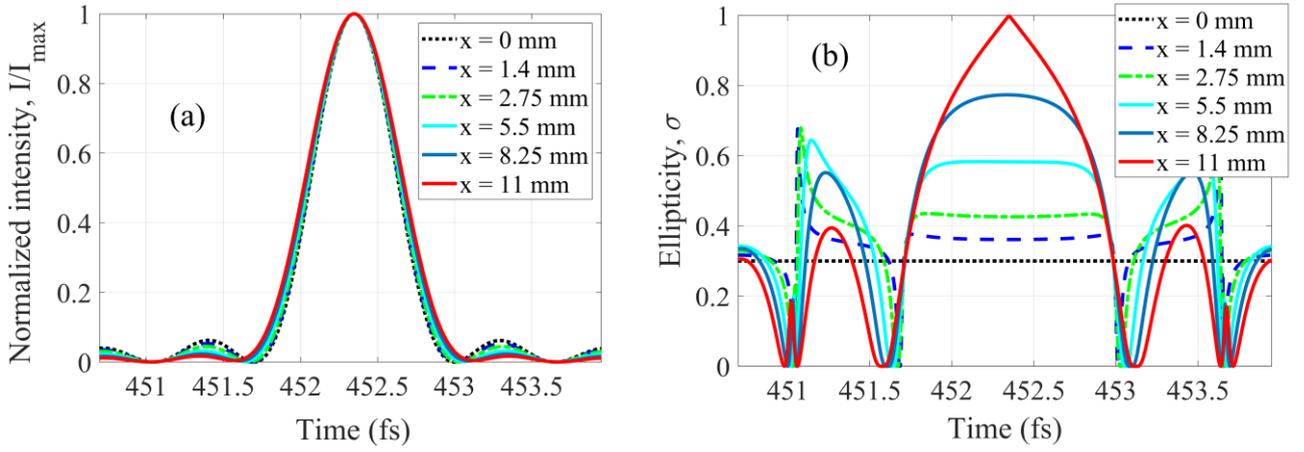

FIG. 17. (a) The shape of individual pulses from the pulse train shown in Fig.16 at different lengths of the medium: $x$=0 mm (incident field), 1.4 mm, 2.75 mm, 5.5 mm, 8.25 mm, and 11 mm. The vertical and horizontal axes show the total intensity of the polarization components of the XUV field and the local time, $\tau$, respectively The time interval corresponds to the peak of the envelope of the total intensity at $x$=11 mm. With increasing length of the medium, the pulse duration grows from $\tau_{pulse}$=590 as at $x$=0 mm to $\tau_{pulse}$=660 as at $x$=11 mm. (b) Variation of ellipticity of the XUV field, corresponding to (a), within a quarter IR-field-cycle. With increasing medium length, the peak ellipticity increases from $\sigma$ = 0.3 at $x$=0 mm to $\sigma$= 0.995 at $x$=11 mm.

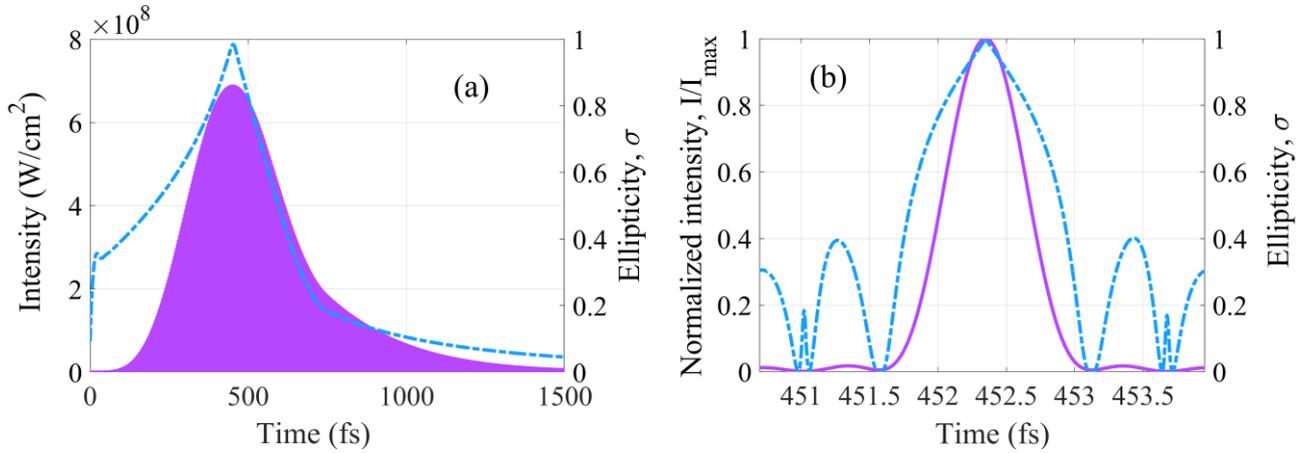

FIG. 18. ((a) Local time dependences of the total intensity (solid lavender curve, left vertical axis) and the ellipticity envelope (dash-dotted blue curve, right vertical axis) of the XUV pulse train, corresponding to Figs. 14-17, amplified in 11 mm long medium. (b) Same as in (a), but within a single half-cycle of the IR field at the maximum of the intensity envelope (in this case, the dash-dotted blue curve shows the instantaneous value of the ellipticity). The XUV intensity in (b) is normalized to its peak value, $I_{max}$=6.9×10$^9$ W/cm$^2$.